\documentclass[reprint,superscriptaddress,amsmath,amssymb,notitlepage,aps,prx,tightenlines]{revtex4-2}

\usepackage[utf8]{inputenc}
\usepackage[caption=false]{subfig}
\usepackage{graphicx}
\usepackage{dcolumn}
\usepackage{bm}
\usepackage{xcolor}
\usepackage{verbatim}
\usepackage{tabularx}
\usepackage{float}
\usepackage{layouts}
\usepackage{hhline}
\usepackage{booktabs}
\usepackage{braket}
\usepackage{comment}
\usepackage{tikz}
\usepackage{quantikz}
\usepackage{upgreek}

\newcommand{\di}{\mathrm{d}}
\newcommand{\bigO}[1]{\mathcal{O}\left( #1 \right)}

\renewcommand{\Re}{\mathrm{Re}}
\renewcommand{\Im}{\mathrm{Im}}

\usepackage[colorlinks=True, citecolor=blue, urlcolor=blue, linkcolor=blue]{hyperref}

\renewcommand{\selectlanguage}[1]{}

\begin{document}

\title{Hardware-efficient quantum phase estimation via local control}

\author{Benjamin F.~Schiffer}
\email[Corresponding author: ]{Benjamin.Schiffer@mpq.mpg.de}
\author{Dominik S.~Wild}
\affiliation{Max-Planck-Institut f\"ur Quantenoptik, Hans-Kopfermann-Str.~1, D-85748~Garching, Germany}%
\affiliation{Munich Center for Quantum Science and Technology (MCQST), Schellingstr.~4, D-80799~Munich, Germany}%

\author{Nishad Maskara}
\author{Mikhail D.~Lukin}
\affiliation{Department of Physics, Harvard University, Cambridge, MA 02138, USA}%

\author{J.~Ignacio Cirac}
\affiliation{Max-Planck-Institut f\"ur Quantenoptik, Hans-Kopfermann-Str.~1, D-85748~Garching, Germany}%
\affiliation{Munich Center for Quantum Science and Technology (MCQST), Schellingstr.~4, D-80799~Munich, Germany}%

\date{\today}

\begin{abstract}
Quantum phase estimation plays a central role in quantum simulation as it enables the study of spectral properties of many-body quantum systems.
Most variants of the phase estimation algorithm require the application of the global unitary evolution conditioned on the state of one or more auxiliary qubits, posing a significant challenge for current quantum devices.
In this work, we present an approach to quantum phase estimation that uses only locally controlled operations, resulting in a significantly reduced circuit depth. 
At the heart of our approach are efficient routines to measure the complex phase of the expectation value of the time-evolution operator, the so-called Loschmidt echo, for both circuit dynamics and Hamiltonian dynamics. 
By tracking changes in the phase during the dynamics, the routines trade circuit depth for an increased sampling cost and classical postprocessing.
Our approach does not rely on reference states and is applicable to any efficiently preparable state, regardless of its correlations.
We provide a comprehensive analysis of the sample complexity and illustrate the results with numerical simulations. 
Our methods offer a practical pathway for measuring spectral properties in large many-body quantum systems using current quantum devices.
\end{abstract}

\maketitle

\section{Introduction}

Quantum simulation is a promising application of quantum computing, enabling the study of many-body quantum systems beyond classical computational capabilities~\cite{feynman_simulating_1982, lloyd_universal_1996, cirac_goals_2012, georgescu_quantum_2014, gross_quantum_2017, altman_quantum_2021, daley_practical_2022}. 
Efficiently simulating quantum processes on quantum hardware is expected to drive advances in condensed matter physics, quantum chemistry, and materials science~\cite{cao_quantum_2019, mcardle_quantum_2020}, providing insights into strongly correlated materials, molecular structures, and non-equilibrium dynamics. 
A widely considered approach is quantum phase estimation (QPE), which enables precise estimation of eigenvalues and serves as a foundation for quantum algorithms in simulation and metrology~\cite{kitaev_quantum_1995, nielsen_quantum_2016, obrien_quantum_2019}. 

\begin{figure}[b]
    \centering
    \includegraphics[width=1.0\linewidth]{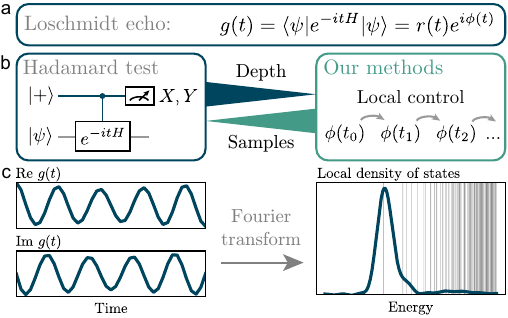}
    \caption{\textbf{(a)} The complex Loschmidt echo quantifies the overlap of an initial state $\ket{\psi}$ with the state obtained after a unitary evolution.
    The evolution can be generated by a Hamiltonian $H$, as illustrated here, or by a quantum circuit.
    \textbf{(b)} While the standard Hadamard test requires global control to measure $g(t)$, our methods extract its phase $\phi(t)$ using only local control, thereby reducing circuit depth at the expense of increased sampling.
    \textbf{(c)} 
    By applying a Fourier transform to the time series of the Loschmidt echo, we obtain a broadened local density of states, shown in front of gray spectral lines.
    }
    \label{fig:Fig1}
\end{figure}

\setlength{\belowcaptionskip}{-6.5pt}
\begin{figure*}[t]
    \centering
    \includegraphics[width=1.0\linewidth]{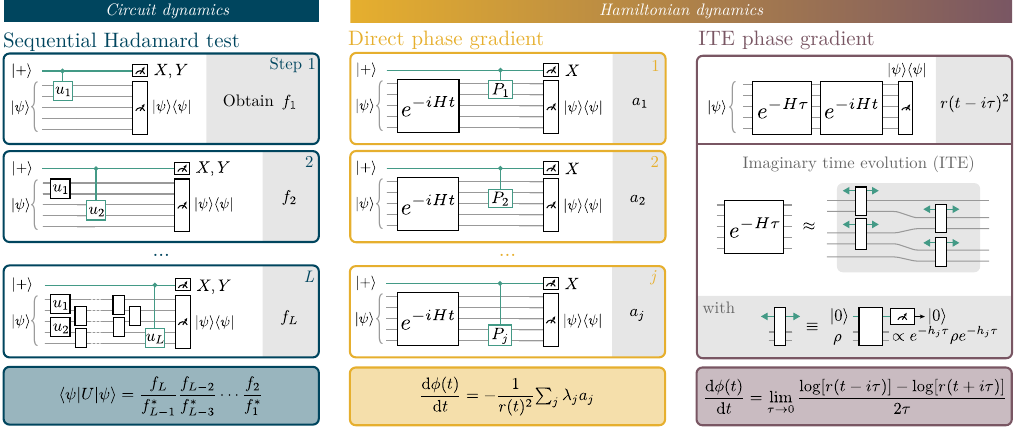}
    \caption{Overview of three methods to obtain the complex Loschmidt echo without global control. 
        \textbf{(Left)} The sequential control method relies on implementing as many different circuits as there are gates, where a single gate is controlled in each circuit.
        It works for arbitrary circuits and is well suited for digital devices.
        \textbf{(Center)} The direct phase gradient method applies to Hamiltonian dynamics, which makes it compatible with hybrid analog-digital devices.
        The method separately computes the derivative of the phase $\di\phi(t)/\di t$ for each local Pauli term $P_j$ of a Hamiltonian $H=\sum_jh_j=\sum_j\lambda_jP_j$. 
        \textbf{(Right)} Alternatively, we can obtain the phase gradient by applying imaginary time evolution (ITE) to an arbitrary initial state using auxiliary qubits. 
        The ITE is realized for each Hamiltonian term $h_j$ separately, using 
        a single auxiliary qubit per term.
    }
    \label{fig:Fig2}
\end{figure*}

The QPE algorithm relies on two key components. 
First, one must prepare a quantum state with significant overlap with the eigenstates of interest, which depends on the specific application.
Second, the phase information must be extracted. 
To obtain the phase, both textbook QPE and more hardware-efficient alternatives such as statistical QPE or virtual filtering~\cite{somma_quantum_2019, ge_faster_2019, lu_algorithms_2021, lin_heisenberg-limited_2022, dutkiewicz_heisenberg-limited_2022, blunt_statistical_2023} require globally controlled operations. 
Concretely, the latter approaches rely on measuring the complex expectation values of the time-evolution operator, which can be obtained using a Hadamard test~[Fig.~\ref{fig:Fig1}(a,b)].
The Hadamard test proceeds by conditioning the evolution of the system on the state of an auxiliary qubit, which incurs substantial cost in terms of the circuit depth. 
As a result, implementing quantum phase estimation remains highly challenging on current devices.

In this work, we introduce three methods to measure the expectation value of a unitary operation for entangled quantum states without requiring global control operations~[Fig.~\ref{fig:Fig1}(b)]. 
The first method applies to arbitrary unitaries defined in terms of a quantum circuit, while the latter two methods concern Hamiltonian dynamics, which are particularly relevant for quantum simulation.
All three methods monitor the change in phase as a function of the evolution time, which can be achieved using only local control or no controlled gates at all.
The algorithms return a time series of the phase, which can be followed by classical post-processing to probe spectral properties of the quantum system, as prescribed by, for instance, statistical QPE or virtual filtering~[Fig.~\ref{fig:Fig1}(c)].

Our approach overcomes several limitations of other methods that avoid the controlled application of the full time-evolution operator.
One such method uses a set of controlled Pauli operators to conditionally flip the sign of the Hamiltonian in a fashion similar to decoupling techniques~\cite{dong_ground-state_2022, chowdhury_controlization_2024}. 
While the controlled Pauli operators may be easier to implement than the conditional evolution,  it is still necessary that they act on the entire system.
This challenge can be avoided using control-free strategies that rely on interference with a reference state for which the value of the Loschmidt echo is known~\cite{lu_algorithms_2021, polla_optimizing_2023, schuckert_probing_2023, hemery_measuring_2024, maskara_programmable_2025}.
However, such reference states are not generally available and even if they are, the preparation of a superposition with the state of interest may be challenging.
Simpler schemes exist for the special case of product states. These include a sequential interferometric protocol that only requires the creation of local superpositions~\cite{lu_algorithms_2021, yang_phase-sensitive_2024} and a method based on short imaginary time evolution (ITE)~\cite{yang_phase-sensitive_2024}, although the utility of these methods is restricted by the fact that product states may not have a significant overlap with the eigenstate of interest.
A different avenue for control-free estimation of Loschmidt echoes was recently taken in Ref.~\cite{clinton_quantum_2024}, where the authors employ classical phase reconstruction techniques.
While the approach may be powerful in practice, it currently lacks a full characterization of its sample complexity.
Our methods require no reference states nor assumptions on the initial state and we provide rigorous bounds on the sample complexity.

The paper is structured as follows. 
In Sec.~\ref{sec:results}, we present a summary of our main results, consisting of three methods for measuring the expectation value of a unitary operator. The methods are described in detail in Sec.~\ref{sec:entangled}, where we provide upper bounds on the sample complexity, discuss practical considerations, and describe simplifications for special cases.
In Sec.~\ref{sec:numerics}, we present a numerical case study of the non-integrable quantum Ising model with an entangled initial state. 
In the simulations, we account for sampling errors and the mitigation of errors due to hardware noise.
We conclude by discussing the broader implications of our work in Sec.~\ref{sec:discussion}.

\section{Summary of results} \label{sec:results}
We present our main result, a family of three methods for estimating the complex expectation value of a unitary operation without global control. 
The first method, which we refer to as the \hyperref[ssec:SequControl]{sequential Hadamard test}, applies to arbitrary quantum circuits composed of $N_\text{gates}$ local gates. The method proceeds by executing shorter circuits involving $1 \leq L \leq N_\text{gates}$ gates, where only a single gate is controlled in each circuit. By sampling from each circuit, it is possible to estimate the Loschmidt echoes
\begin{align} \label{eq:loschmidt_circuit}
    g_L=r_Le^{i\phi_L}=\langle\psi|U_L|\psi\rangle,
\end{align}
where $U_L$ is the unitary corresponding to the first $L$ gates.

The second and third methods (\hyperref[ssec:DirectGrad]{direct phase gradient} and \hyperref[ssec:BE-ITE]{ITE phase gradient}) concern the Loschmidt echo associated with the unitary $U(t) = \exp(-i H t)$, generated by a Hamiltonian $H$. 
We assume that the Hamiltonian is a sum of local terms, which, without loss of generality, may be expressed as $H = \sum_{j=1}^K \lambda_j P_j$. Here, each $P_j$ is a Pauli operator acting nontrivially on a constant number of qubits and $\lambda_j$ is a real coefficient bounded in magnitude by a constant. We further assume that there are $K = \bigO{n}$ terms in the sum, as is the case for any finite-ranged interaction on a lattice in $d$ dimensions.
The Loschmidt echo is a continuous function of time, defined by
\begin{align}
    g(t)=r(t)e^{i\phi(t)}=\langle\psi|U(t)|\psi\rangle.    
\end{align}
The two methods rely on the insight that the derivative of the phase, $\di \phi(t) / \di t$, can be measured without global control. In the direct phase gradient method, we accomplish this by controlled application of individual local Hamiltonian terms. In the ITE phase gradient method, we instead apply a short imaginary-time evolution (ITE). In contrast to previous work~\cite{yang_phase-sensitive_2024}, we realize the ITE using local block encodings, which allows us to extend the approach beyond product states to arbitrary initial states. Both methods additionally require unconditional evolution under the Hamiltonian, which may be implemented using analog or digital quantum simulation techniques.

A visual overview of our methods is presented in Fig.~\ref{fig:Fig2}.
We note that all three methods can be readily generalized to compute the expectation value of products of unitaries~\cite{yang_phase-sensitive_2024}, which also provides access to quantities of the form $\langle\psi'| U |\psi\rangle$ for different states $\ket{\psi}$ and $\ket{\psi'}$. The generalized form allows to compute the expectation values of observables of eigenstates, in addition to the eigenenergies obtained from phase estimation~\cite{lu_algorithms_2021}.
Below, we describe the methods in detail and analyze the sampling cost to compute the phase $\phi_L$ or $\phi(t)$ within additive error $\epsilon$. 
To compare the methods, we consider a product formula approximation to the Hamiltonian evolution for the sequential Hadamard test.
The results are summarized in Table~\ref{tab:methods_entangled}. 
Each method yields not only the final phase but also its value at intermediate times. The intermediate times depend on the method, but in each case the number of points grows at least linearly with $t$.

As the comparison shows, the sequential Hadamard test is competitive with the two Hamiltonian-specific approaches, and its broad applicability makes it the preferred method for circuit-based simulations.
For analog quantum simulation, one may choose between the two Hamiltonian‐specific methods. If auxiliary qubits are scarce, the direct phase gradient method provides the simplest and most resource‐efficient implementation. 
When a sufficient number of auxiliary qubits is available, the ITE phase gradient offers an alternative with a better dependence on the magnitude of the Loschmidt echo, although  at the expense of a slightly less favorable dependence on $t$ and $\epsilon$.

For sufficiently small errors, the uncertainty in the Loschmidt echo $g$ is related to the uncertainty in the phase $\phi$ by $|\Delta g|^2 = (\Delta r)^2 + (r\Delta\phi)^2$, where the amplitude $r$ can be estimated with uncertainty $\Delta r \leq \epsilon$ using $\bigO{1/\epsilon^2}$ samples (see Appendix~\ref{app:project}).
Therefore, the sample complexity required to compute the phase with additive error $\epsilon$ is also sufficient to compute the full Loschmidt echo $g$ within the same error.
For comparison, determining the phase with error $\epsilon$ using the Hadamard test requires $\bigO{1/ (\epsilon r)^2}$ samples. 
To obtain the phase for $\bigO{t}$ time points, the sample complexity is $\bigO{t / (\epsilon r_\text{min})^{2}}$, where $r_\text{min} = \min_{t' \in [0, t]} \{ r(t') \}$. 
The dependence on $r_\text{min}$ arises because estimating the phase of a small quantity requires resolving its magnitude. 
The Hadamard test avoids this dependence when the goal is to estimate $g$ with additive error $\epsilon$, for which $\bigO{t / \epsilon^2}$ samples are sufficient.
The dependence on $r_\text{min}$ is unavoidable with our methods even when estimating $g$ as the methods proceed by tracking the phase throughout the evolution.
% If the amplitude revives after first vanishing, combining the methods presented here with the Hadamard test can mitigate the sampling overhead incurred by small intermediate amplitudes.

\begin{table}[t!]
    \renewcommand{\arraystretch}{2.3}
    \setlength{\tabcolsep}{6.5pt}
    \centering
    \begin{tabular}{c|c}
        \textbf{Method}  & \textbf{Sample complexity} \\
        \hhline{=|=}
        \hyperref[ssec:SequControl]{Seq. Hadamard test} & $\bigO{(n t / \epsilon)^{2 + 2/p}r_\text{min}^{-2}}$ \\
        \hline
        \hyperref[ssec:DirectGrad]{Direct phase gradient} & $\bigO{(n t / \epsilon)^{2 + 1/s}r_\text{min}^{-4}}$ \\
        \hline
        \hyperref[ssec:BE-ITE]{ITE phase gradient} &  $\bigO{n^{2 + 1/s} (t / \epsilon)^{3 + 3/2s} r_{\pm, \text{min}}^{-2}}$ \\
        \hhline{=|=}
    \end{tabular}
    \caption{Overview of methods for estimating the phase of a Loschmidt echo  without global control operations. 
    We consider the total sample complexity to compute the phase of the Loschmidt echo at time $t$ for a system of $n$ qubits within additive error $\epsilon$.
    Phases are also computed at intermediate times, which depend on the method.
    The quantity $r_\text{min}$ denotes the smallest amplitude encountered during the evolution, i.e., $r_\text{min} = \min_{l \in [1, N_\text{gates}]} \{r_l\}$ for the sequential Hadamard test and $r_\text{min} = \min_{t' \in [0, t]} \{r(t')\}$ for the direct phase gradient.
    The ITE phase gradient includes an imaginary time evolution for short~$\tau$, with which $r_{\pm,\min} = \min_{t' \in [0, t]}\{ r(t'+i\tau),r(t'-i\tau) \}$.
    The integers $p$ and $s$ denote respectively the order of the product formula and the numerical integration method ($s=2$ for Simpson's rule).
    }
    \label{tab:methods_entangled}
\end{table}

\section{Local measurement schemes} \label{sec:entangled}
\subsection{Sequential Hadamard test} \label{ssec:SequControl}

Our first method can be applied to any unitary of the form
\begin{align}
U = u_{N_\text{gates}} \cdots u_2 u_1,
\end{align}
where each $u_l$ is a local gate. In the context of quantum simulation, the circuit may approximate the evolution under a Hamiltonian, although this is not necessary. For instance, the method also applies to Floquet dynamics with large time steps, where an explicit Hamiltonian representation may not be available~\cite{yang_simulating_2023, kim_evidence_2023}.
The method is also compatible with other methods of digital quantum simulation, including randomized compiled circuits, which may be advantageous for quantum chemistry applications~\cite{campbell_random_2019, wan_randomized_2022, gunther_phase_2025}.
To minimize the circuit depth, fermion-to-qubit mappings that preserve locality may further be useful in this context~\cite{bravyi_fermionic_2002, verstraete_mapping_2005}.

The sequential Hadamard test proceeds iteratively.
We begin by applying only the first Trotter gate $u_1$, controlled by an auxiliary qubit. 
Then, the first Trotter gate $u_1$ is applied without control, followed by the second Trotter gate $u_2$ with control. 
Next, the third Trotter gate is controlled, and so forth. 
The complex Loschmidt echo is obtained recursively from the information of each previous step as summarized in the left panel of Fig.~\ref{fig:Fig2}.

\begin{figure}[t]
    \centering
    \begin{quantikz}
        \lstick{$\ket{+}$} & \qw & \ctrl{1} &  \meter{\{\sigma^x, \sigma^y\}} \\
        \lstick{$\ket{\psi}$} & \gate{\prod_{j=1}^{L-1} u_j} & \gate{u_L}  & \meter{|\psi\rangle\langle\psi|}
    \end{quantikz}
    \caption{Circuit diagram for the $L^\text{th}$ iteration of the sequential Hadamard test.}
    \label{fig:sequ_grad_1}
\end{figure}
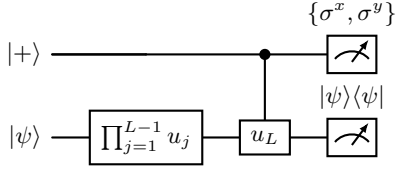

More explicitly, the $L^\text{th}$ iteration consists of repeated measurements of the output of the circuit shown in Fig.~\ref{fig:sequ_grad_1}. 
Projecting the state register onto the initial state and measuring the auxiliary qubit yields~\footnote{The expectation value $\langle \sigma^x + i \sigma^y\rangle$ can also be evaluated directly using two controlled operations.}
\begin{align}
    \langle \sigma^x \otimes |\psi\rangle\langle\psi|\rangle = \Re(g_L g_{L-1}^*),\label{eq:sht_re}\\
    \langle \sigma^y \otimes |\psi\rangle\langle\psi|\rangle = \Im(g_L g_{L-1}^*)\label{eq:sht_im},
\end{align}
where $g_L$ is the Loschmidt echo in Eq.~\eqref{eq:loschmidt_circuit} with $U_L = u_L \cdots u_2 u_1$.
The projection onto the initial state can be realized by reversing the preparation of $\ket\psi$ and performing a computational basis measurement (see Appendix~\ref{app:project} for details). Hence, these measurements allow us to reconstruct the Loschmidt echo as
\begin{align}
    g_L=\frac{g_L^{} g_{L-1}^*}{g_{L-1}^*g_{L-2}^{}} \cdot \frac{g_{L-2}^{}g_{L-3}^*}{g_{L-3}^*g_{L-4}^{}} \cdot \ldots \cdot \frac{g_2^{}g_1^*}{g_1^*}.
    \label{eq:recursive}
\end{align}
Here, we used the fact that $g_0 = 1$ and assumed that $L$ is even, although a similar expression holds for odd $L$. We emphasize that each numerator and denominator is measured separately, requiring only local control.

We observe that the complex argument of $g_L^{}g_{L-1}^*$ is equal to the phase difference $\delta\phi_L = \phi_L - \phi_{L-1}$. We may thus also compute the phase of the Loschmidt echo directly as $\phi_L = \sum_{l=1}^L (-1)^{L-l}\,\delta\phi_l$, which is equivalent to taking the complex argument of Eq.~\eqref{eq:recursive}. 
In practice, the finite number of measurements leads to a statistical uncertainty in the values of the observables in Eqs.~\eqref{eq:sht_re} and~\eqref{eq:sht_im}, which in turn results in an uncertainty in the phase difference $\delta \phi_l = \tan^{-1}[\Im(g_l^{}g_{l-1}^*)/\Re(g_l^{}g_{l-1}^*)]$. 
Assuming that we measure each observable $M_l$ times, the variance in the phase difference is given by (see Appendix~\ref{app:ssec:sequ} for details)
\begin{align}\label{eq:sht_cancel}
    \left(\Delta [\delta \phi_l]\right)^2 &\leq \frac{1}{2M_l} \Big( \frac{1}{r_{l-1}^2} + \frac{1}{r_{l}^2} \Big).
\end{align}
Since the samples obtained from separate measurements are statistically independent, the  phase error of the Loschmidt echo is given by $(\Delta \phi_L)^2 \approx \sum_{l=1}^{L} (\Delta [\delta \phi_l])^2$.
We choose the number of measurements for a given value of $L$ as $M_l = L (1/r_{l-1}^2 + 1/r_l^2) / 2\epsilon^2$ such that each term in the sum contributes $\epsilon^2 / L $ and the error in $\phi_L$ is bounded from above by $\epsilon$.
We thus require $M_\mathrm{tot} = \sum_{l=1}^L M_l \leq  (L / \epsilon \,r_\text{min})^2$ samples to determine $\phi_L$ within additive error $\epsilon$, where $r_\text{min}=\min_l\{r_l\}$.
The total sample complexity is thus bounded by
\begin{equation}
    M_\text{tot} = \mathcal{O}\left(N_\text{gates}^2\, \epsilon^{-2}\,r_\text{min}^{-2} \right).
\end{equation}
We highlight that this number of samples is sufficient to not only determine the final phase but also all intermediate phases $\phi_L$ and amplitudes $r_L$ with the same additive error $\epsilon$.

While this approach works for any circuit, we consider the particular application to Hamiltonian dynamics, allowing us to compare the sample complexity with the subsequent two methods (see Table~\ref{tab:methods_entangled}). Given a local Hamiltonian $H$ acting on $n$ qubits, the time evolution operator $\exp(-i H t)$ may be implemented using a $p^\text{th}$ order Suzuki-Trotter decomposition, requiring $N_\text{gates} = \bigO{(n t)^{1 + 1/p} / \epsilon^{1/p}}$ gates~\cite{hatano_finding_2005, childs_theory_2021}. This results in the sample complexity $M_\text{tot} = \bigO{(n t / \epsilon)^{2 + 2/p} r_\text{min}^{-2}}$.

As a generalization of the above approach, it is possible to sequentially add the gates to the circuit in any order. Whenever a new gate is added, it is inserted into the correct location of the circuit and its application is controlled by an auxiliary qubit, while all previously added qubits are applied unconditionally. This freedom can be used to reduce the sample complexity, in particular, when the initial state is a product state.
If we remove all gates of the circuit that entangle two subsystems, we can effectively cut the circuit in half along the time dimension. 
Continuing to cut the remaining system, eventually, we are left with computing the complex Loschmidt echo of a few-qubit system, which can be done on a classical computer.
In particular, for translationally invariant systems, the remaining subsystems can be identical, leading to significant resource savings.
This circuit cutting technique may also be useful for initial states with exponentially decaying correlations, such as a matrix product state with a constant bond dimension, or for states prepared using shallow circuits. 
We refer to the Appendices~\ref{app:cutting} and \ref{app:product_sht} for more details.

When a reference state with known Loschmidt echo is available, it can be combined with the sequential Hadamard test to track the change of the phase information from the reference state to the initial state.
We provide details for this approach in Appendix~\ref{app:ref_state}.
In addition, we show that it is possible to avoid auxiliary qubits entirely to obtain the complex expectation value $\langle \psi|u_LU|\psi\rangle$ where $u_L=e^{i\theta_L P_L}$ is a Pauli rotation and $\langle \psi|U|\psi\rangle$ is known. 
By choosing Pauli matrices such that $P_1P_2=P_L$ and measuring amplitudes $|\langle \psi|e^{i\theta P}U|\psi\rangle|^2$ and $|\langle \psi|e^{i\theta_1 P_1}e^{i\theta_2 P_2}U|\psi\rangle|^2$ with appropriately chosen angles, one obtains a variant that does not require any controlled operations or auxiliary qubits (see Appendix~\ref{app:no_ancilla} for details).

\subsection{Direct phase gradient} \label{ssec:DirectGrad}

The second method pertains to the complex Loschmidt echo 
\begin{align}
    g(t) = \braket{\psi | e^{- i H t} | \psi}.
\end{align}
The method is agnostic to the implementation of the time-evolution operator.
The Hamiltonian dynamics may be realized using any digital quantum simulation method or by means of analog simulation, rendering this approach ideally suited to emerging hybrid analog-digital quantum platforms~\cite{bluvstein_quantum_2022, andersen_thermalization_2025}.

\begin{figure}
    \centering
    \begin{quantikz}
        \lstick{$\ket{+}$} & & \ctrl{1} & \meter{\sigma^x} \\
        \lstick{$\ket{\psi}$} & \gate{e^{-iHt}} & \gate{P_j} & \meter{|\psi\rangle\langle\psi|}
    \end{quantikz}
    \caption{Circuit diagram for the direct phase gradient method. The measurement yields the expression in Eq.~\eqref{eq:direct_phase_circuit}.}
     \label{fig:directGradCircuit}
\end{figure}
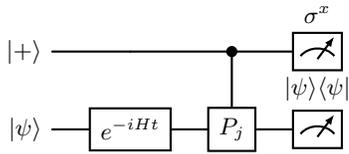

If the Loschmidt echo is nonzero, its complex phase may be expressed as $\phi(t) = \Im[\log g(t)]$, where the branch of the logarithm can be chosen such that $\phi(t)$ is continuous and differentiable. 
Direct differentiation yields
\begin{align}\label{eq:dpg}
    \frac{\di\phi(t)}{\di t} = -\frac{1}{2r(t)^2} \bra{\psi} e^{iHt} \{H, |\psi\rangle\langle\psi|\} e^{-iHt} \ket{\psi},
\end{align}
where $\{A,B\}=AB+BA$ denotes the anti-commutator. Since $\{H, \ket{\psi} \bra{\psi} \}$ is Hermitian, the right-hand side can be viewed as the expectation value of an observable. To measure it using local operations, we employ the local decomposition of the Hamiltonian. In particular, by measuring the expectation value $\langle \sigma^x \otimes |\psi\rangle\langle\psi| \rangle$ at the end of the circuit shown in Fig.~\ref{fig:directGradCircuit}, we obtain
\begin{equation}
    \label{eq:direct_phase_circuit}
    a_j(t) = \frac{1}{2} \bra{\psi(t)} \{P_j, |\psi\rangle\langle\psi|\}  \ket{\psi(t)}.
\end{equation}
By implementing this circuit for each term in the Hamiltonian and separately measuring $r(t)$, we are able to compute the phase gradient $\di \phi(t)/\di t = - \sum_j \lambda_j a_j(t) / r(t)^2$. The phase is finally obtained by numerically integrating the phase gradient using intermediate measurements at small time intervals with the initial condition $\phi(0) = 0$. The complete procedure is summarized in the central panel of Fig.~\ref{fig:Fig2}.

In practice, we can perform measurements only for a discrete set of times $\{t_l\}$ with $l \in \{1, 2, \ldots, N_\text{times} \}$, for which we define the shorthands $\phi'_l = \left. \di \phi(t) / \di t \right|_{t = t_l}$,  $a_{j,l} = a_j(t_l)$, and $p_l = r(t_l)^2$. 
The variance of $\phi'_l$ can be expressed in terms of the variance of the measured quantities as
\begin{equation}
    \left(\Delta \phi'_l\right)^2 \approx \frac{1}{p_l^2} \sum_{j=1}^K \lambda_j^2 \Delta a_{j,l}^2 + \left( \phi'_l \right)^2 \frac{\Delta p_l^2}{p_l^2} .
\end{equation}
The uncertainties $\Delta a_{j,l}$ and $\Delta p_l$ depend on the number of measurements $M^{(a)}_{j,l}$ and $M^{(r)}_l$ used to estimate $a_{j,l}$ and $p_l$, respectively.
As shown in Appendix~\ref{app:ssec:dpg}, the number of measurements $M_l = \sum_{j = 1}^K M^{(a)}_{j,l} + M^{(r)}_l = \bigO{(n / \eta p_l)^2}$ in time step $l$ is sufficient to compute the phase gradient $\phi_l'$ with additive error $\eta$.

In addition to the statistical uncertainty in the phase gradient, the estimation of the final phase incurs a numerical integration error. Denoting the integration error by $\Delta \phi_I$, the total error satisfies $\Delta \phi \leq N_\text{times}^{1/2} \eta + \Delta \phi_I$, where we chose the same statistical error $\eta$ at each time step and used the fact that the statistical errors are uncorrelated.
For the Newton-Cotes formula~\footnote{In practice, other numerical integration methods with uneven spacing, such as those based on Gaussian quadrature rules, could yield better integration results, but bounding the integration error is more challenging.}, the integration error is $\Delta\phi_I = \bigO{n t  (t/N_\text{times})^{2 s}}$, where $s$ is an integer that depends on the order of the formula ($s = 1$ for the trapezoidal rule, $s = 2$ for Simpson's rule)~\cite{quarteroni_numerical_2007}. By choosing $N_\text{times} = \bigO{(n t / \epsilon)^{1/2s} \, t}$ and $\eta = \bigO{\epsilon / N_\text{times}^{1/2}}$, we find
that $\Delta \phi \leq \epsilon$ with the total number of measurements bounded by
\begin{equation}
    M_\mathrm{tot} = \sum_{l = 1}^{N_\text{times}} M_l = \bigO{(n t / \epsilon)^{2 + 1/s} r_\text{min}^{-4}},
\end{equation}
where $r_\text{min}=\min_{t' \in [0, t]}\{r(t')\}$.
We compare this result with the sample complexity of the other methods in Table~\ref{tab:methods_entangled}. 

The direct phase gradient method can be used without auxiliary qubits if the states $\ket{\varphi_j^\pm} = (\ket{\psi} \pm P_j \ket{\psi})/\sqrt{2}$ can be efficiently prepared. This is the case, for instance, if $\ket{\psi}$ is a product state. The quantities $a_j$ can be obtained from $2 a_j = |\braket{\psi | \exp(-iHt) | \phi_j^+} | ^2 - |\braket{\psi | \exp(-iHt) | \phi_j^-} | ^2$, where the two terms on the right-hand side can be measured by evolving $\ket{\phi_k^\pm}$ and performing a projective measurement onto $\ket{\psi}$.
Instead of decomposing the Hamiltonian into local Pauli terms, it can be beneficial in some cases to prepare the states
$\ket{\varphi^\pm} = (\ket{\psi} \pm \ket{\psi_H})/\sqrt{2}$, where
$\ket{\psi_H}=H|{\psi}\rangle\,/\,||H|\psi\rangle||$.
While this can generally be achieved with global block-encoding techniques~\cite{gilyen_quantum_2019, rall_quantum_2020}, these techniques are highly demanding in practice. 
For some special cases, however, there are relatively simple ways to prepare $\ket{\varphi^\pm}$. 
In particular, for a geometrically-local, one-dimensional system, these states can be written as a matrix product state of constant bond dimension. 
Then, the states $\ket{\varphi^\pm}$ can be prepared exactly with a sequential quantum circuit in $\bigO{n}$ depth~\cite{schon_sequential_2005}.
This allows to effectively reduce the sample complexity by a factor of $\bigO{n^2}$ at the expense of a slightly deeper circuit.

The direct phase gradient method can also be realized without a control qubit, similarly to the sequential Hadamard test. 
This allows to obtain the phase of $\langle \psi|P_jU(t)|\psi\rangle$ using only amplitude measurements, from which we can compute $a_j(t)$ with Eq.~\eqref{eq:direct_phase_circuit}.
We include the details on this control-free variant in Appendix~\ref{app:no_ancilla}.

\subsection{ITE phase gradient} \label{ssec:BE-ITE}

Our third method also applies to the continuous Loschmidt echo $g(t)$. Instead of local control of the Hamiltonian terms, it uses a short imaginary time evolution (ITE) to measure the phase gradient. This approach was originally proposed in Ref.~\cite{yang_phase-sensitive_2024}, where the initial states were, however, restricted to states with only short-ranged correlations. We generalize this result to arbitrary initial states by realizing the ITE with local block-encodings. Like the direct phase gradient, this method is independent of the implementation of the real-time dynamics and is thus suitable for hybrid analog-digital devices. 

The method relies on promoting the time $t$ to a complex variable $z = t - i \beta$. If the Loschmidt echo does not vanish, then $\log g(z) = \log r(z) + i \phi(z)$ is an analytic function for any finite-sized system. The Cauchy-Riemann equations then allow us to relate the phase gradient to the derivative of $\log r(z)$ with respect to imaginary time according to
\begin{align} \label{eq:ite}
    \frac{\di \phi(t)}{\di t} &= \left. \frac{\partial \log r(t - i \beta)}{\partial \beta}\right|_{\beta=0}.
\end{align}
We will approximate the right-hand side by the finite difference $[\log r(t - i \tau) - \log r(t + i \tau)] / 2 \tau$ for a small imaginary time step $\tau$. Hence, the phase gradient can be computed from the quantities $r(t \pm i \tau)^2 = |\langle{\psi}|e^{-i H t}e^{\pm H \tau}|{\psi}\rangle|^2$, which correspond to survival probabilities following a short initial ITE.

It was shown in Ref.~\cite{yang_phase-sensitive_2024} that if $\ket{\psi}$ is a product state, the state $e^{\pm H \tau}|{\psi}\rangle$ can be approximately prepared (up to normalization) using a constant-depth unitary circuit. For entangled states with longer-ranged correlations, however, this approach becomes intractable.
Here, we instead implement $e^{\pm H \tau}$ by acting jointly on the system and a set of auxiliary qubits before postselecting a specific outcome of a measurement of the auxiliary qubits.
The postselection probability can be made large because the imaginary time $\tau$ is small. The key steps of the ITE phase gradient method are summarized in the right panel of Fig.~\ref{fig:Fig2}.

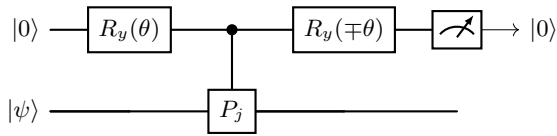
\begin{figure}[t]
    \centering
    \begin{tikzpicture}
    \node[scale=1]{
    \begin{quantikz}
        \lstick{$\ket{0}$} & \gate{R_y(\theta)} & \ctrl{1} & \gate{R_y(\mp\theta)} &  \meter{} \arrow[r]  & \rstick{$\ket{0}$}\setwiretype{n} \\
        \lstick{$\ket{\psi}$} & \qw & \gate{P_j} & \qw & 
    \end{quantikz}
    };
    \end{tikzpicture}%
    \caption{Circuit used to apply the local ITE operator $\propto\exp(\pm  P_j\lambda_j\tau)$ to an initial quantum state $\ket{\psi}$ using $R_y=\exp(-i\theta\sigma^y/2)$ rotations on the single auxiliary qubit, where $\theta$ depends on $\lambda_j$ (see main text).
    The ITE is successfully applied when the auxiliary qubit is measured to be in the $\ket{0}$-state at the end of the circuit.}
    \label{fig:LocalBE}
\end{figure}

To realize the ITE, we use the first-order Trotter decomposition $e^{\pm H \tau} \approx \prod_j e^{\pm P_j\lambda_j\tau}$.
Each term in the product is implemented using the circuit in Fig.~\ref{fig:LocalBE}, which may be viewed as a special case of a block encoding~\cite{gilyen_quantum_2019, rall_quantum_2020} and as a linear combination of unitaries in particular~\cite{childs_hamiltonian_2012}.
After postselecting on the auxiliary qubit being in the $\ket{0}$ state, the circuit prepares the (unnormalized) state
\begin{align} \label{eq:ITE_cos}
    [\cos^2(\theta/2)\mathbb{I} \pm \sin^2(\theta/2)P_j] \ket{\psi}.
\end{align}
By comparing this expression with the identity
\begin{equation}
    e^{\pm P_j\lambda_j\tau} = \cosh(\lambda_j\tau ) \mathbb{I} \pm \sinh(\lambda_j\tau ) P_j,
\end{equation}
we can see that the postselected circuit thus implements the non-unitary operation $e^{-  |\lambda_j|\tau} e^{\pm P_j\lambda_j\tau}$ if we choose $\theta$ such that $\cos^2 (\theta / 2) =  e^{-|\lambda_j|\tau} \cosh(\lambda_j\tau)$ and $\sin^2 (\theta / 2) =  e^{- |\lambda_j|\tau} \sinh(|\lambda_j|\tau)$.
The prefactor $e^{-  |\lambda_j|\tau}$ is accounted for in classical postprocessing by simply rescaling the data.
The success probability of each measurement is given by the norm squared of the state in Eq.~\eqref{eq:ITE_cos}. For small values of $\tau$, the probability that all measurements succeed is thus given by
\begin{align}
    P_\mathrm{succ} = 1 - 2 \tau \sum_j \left( |\lambda_j| \mp \lambda_j \braket{\psi | P_j | \psi} \right) + \bigO{\tau^2}.
\end{align}
Since there are $\bigO{n}$ Hamiltonian terms, the failure probability for a short ITE is $P_\text{fail} = \bigO{n\tau}$. If any one of the auxiliary qubits heralds failure, the protocol is restarted.
To achieve a constant success probability, independent of system size, we require that the imaginary-time duration decreases with system size, $\tau = \bigO{1/n}$. 

As for the previous method, shot noise leads to a statistical uncertainty in the phase gradient $\phi'_l = \left. \di \phi(t) / \di t \right|_{t = t_l}$. With the shorthand $p_{l}^\pm = r(t_l\pm i\tau)^2$, the expected variance of $\phi'_l$ is given by
\begin{align}
    (\Delta \phi'_l)^2 \approx \frac{1}{(4 \tau)^2} \left[ \left(\frac{\Delta p_l^+}{p_l^+} \right)^2 + \left( \frac{\Delta p_l^-}{p_l^-} \right)^2 \right] .
\end{align}
Using the above implementation of the imaginary time evolution, we obtain $p_l^\pm$ by performing a projective measurement onto $\ket{\psi}$. The corresponding uncertainty satisfies 
$(\Delta p_l^\pm)^2 \leq p_l^\pm / M_l^\pm$, where $M_l^\pm$ denotes the respective number of measurements. 
Hence, $M_{l}^\pm=\bigO{1/(\tau^2\eta^2 p_{l}^\pm)}$ samples are sufficient to ensure that the error $\Delta \phi'_l$ due to shot noise is bounded from above by $\eta$. The total number of samples is given by $M_\text{tot} = \sum_{l=1}^{N_\text{times}} (M_l^+ + M_l^-)= \bigO{(N_{\text{times}} / \tau \epsilon r_{\pm,\min})^2}$, where $r_{\pm, \text{min}} = \min_{t' \in [0, t]} \{ r(t' \pm i \tau) \}$.

Following the discussion in the previous section~\ref{ssec:DirectGrad}, 
we numerically integrate the phase gradient and choose $\eta = \bigO{\epsilon / N_\text{times}^{1/2}}$ to bound the accumulated statistical uncertainty by~$\epsilon$. 
In addition to the statistical uncertainty and numerical integration error, the ITE phase gradient incurs a systematic error due to the first-order Trotter decomposition of the imaginary time evolution and the finite-difference approximation of the derivative in Eq.~\eqref{eq:ite}. 
Assuming that the Loschmidt echo and a closely related quantity do not vanish, both errors can be shown to be of size $\bigO{n \tau^2}$~\cite{yang_phase-sensitive_2024}, which may accumulate during the numerical integration to $\bigO{N_\text{times} n \tau^2}$. 
By choosing $\tau = \bigO{(\epsilon / t)^{(1 + 1/2s)/2} / n}$, we ensure that this total error bounded by $\epsilon$ while satisfying the above requirement $\tau = \bigO{1/n}$. 
By plugging into the expression for $M_\text{tot}$, we finally obtain
\begin{equation}
    M_\text{tot} = \bigO{n^{2 + 1/s} (t / \epsilon)^{3 + 3/2s} r_{\pm, \text{min}}^{-2}}.
\end{equation}
As can be seen from the comparison in Table~\ref{tab:methods_entangled}, the ITE phase gradient compares favorably to the direct phase gradient in terms of the dependence on the magnitude of the Loschmidt echo, although the dependence on $t$ and $\epsilon$ is slightly worse.

We note that our approach is compatible with other methods of implementing ITE, such as double-bracket algorithms~\cite{gluza_double-bracket_2024}.
For product states, short ITE can be implemented without any auxiliary qubits, as proposed in~\cite{yang_phase-sensitive_2024}.  
We further remark that a simplified implementation is possible when the Hamiltonian can be split into two as $H = H_1 + H_2$ and the initial state is an eigenstate of $H_1$ with eigenvalue $\zeta_1$. 
Then, at a first-order Trotter decomposition, the ITE only yields a global prefactor $\exp(-\zeta_1\tau)$:
\begin{align}
    e^{-H \tau} \ket{\psi} \approx e^{-H_2\tau } e^{-H_1\tau } \ket{\psi} = e^{-\zeta_1\tau } e^{-H_2\tau } \ket{\psi}.
\end{align}
This simplification is particularly interesting when the Hamiltonian can be split such that $H_2$ only contains single-qubit operators.
This is the case for Ising-type Hamiltonians, including the Hamiltonian describing neutral atoms interacting via Rydberg interactions, as discussed in detail in Appendix~\ref{app:Rydberg}. Similar simplifications apply to the direct phase gradient method since the contribution from $H_1$ in Eq.~\eqref{eq:dpg} is $- \zeta_1$. It thus suffices to measure $a_j$ for the terms appearing in $H_2$.

\begin{figure}[t]
    \centering
    \includegraphics[width=1.0\linewidth]{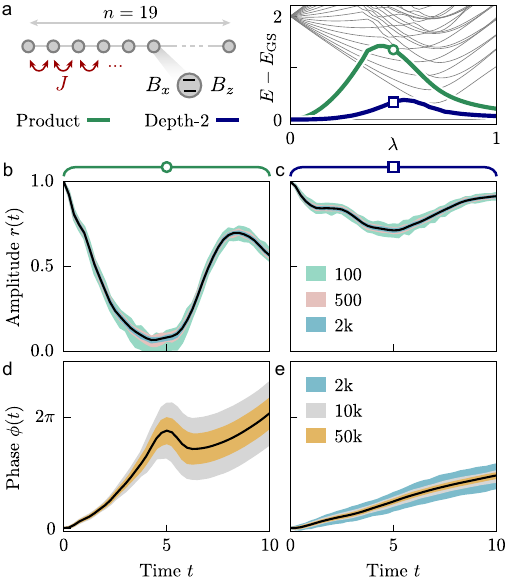}
    \caption{\textbf{(a)} Numerical case study of the sequential Hadamard test for a one-dimensional Ising model with a longitudinal ($B_z$) and transverse field ($B_x$) on 19 spins. 
    For the family of Hamiltonians $H(\lambda)$ introduced in the main text, the gray lines depict the energy differences $E-E_\text{GS}$ for the low-energy states above the ground state.
    We include the lowest achievable energies for both a product state ansatz and a depth-2 variational ansatz. 
    The following panels apply to the point $\lambda=0.5$ in vicinity of the critical point, highlighted by markers.
    We show the exact amplitude $r(t)$ of the Loschmidt echo as a line in \textbf{(b)} for the product state and in \textbf{(c)} for the entangled state.
    Shaded areas are estimated with a fixed shot number per measurement---as specified in the legend---and indicate one standard deviation above and below the mean, estimated with 50 independent data sets. 
    The phase $\phi(t)$, reconstructed using the sequential Hadamard test, is shown in \textbf{(d,e)} for the two states.
    The contributions of the respective mean energies to the phase have been subtracted for clarity.
    }
    \label{fig:Fig3}
\end{figure}

\section{Numerical case study} \label{sec:numerics}

In this section, we numerically showcase the sequential Hadamard test for a simple spin model, including sampling and hardware noise.
We consider the non-integrable, one-dimensional Ising model with both a longitudinal and a transverse field~[Fig.~\ref{fig:Fig3}(a)]:
\begin{align}
    H_\text{Ising}=J \sum_i \sigma_i^z \sigma_{i+1}^{z} + B_x \sum_i\sigma_i^x + B_z \sum_i \sigma_i^z.
\end{align}
We parametrize a family of Hamiltonians by $\lambda$, $H(\lambda)=(1-\lambda)H_0 + \lambda H_1$, which interpolates between $H_0 = H_\text{Ising}(J\,=\,0, B_x\,=\,1, B_z\,=\,0)$ and  $H_1 = H_\text{Ising}(1.5, 1, 1)$. 
The spectrum of $H(\lambda)$ is shown in Fig.~\ref{fig:Fig3}(a), along with curves indicating the optimized energies using two probe states: a product state, and a depth-2 variational circuit. 
Details of the ansätze are included in Appendix~\ref{app:sec:Ising}.
Although the product state ansatz attains relatively low energies in this Ising model, in general, preparing an entangled initial state is crucial for accessing lower-energy states. 
We choose the Hamiltonian $H(\lambda=0.5)$ and simulate the sequential Hadamard test using both optimized initial states.

The time evolution is implemented using a second-order Suzuki-Trotter decomposition of Hamiltonian evolution $\exp(-iHt)$ with a Trotter step of $0.25$.
We show the amplitude $r(t)$ and the reconstructed phase $\phi(t)$ of the complex Loschmidt echo for both initial states in Fig.~\ref{fig:Fig3}(b-e).
While the amplitude can be estimated without control, the phase information is obtained using the sequential Hadamard test. 
In the absence of sampling noise, our method reconstructs the exact phase for the Trotterized circuit.

To quantify the influence of sampling noise, we simulate the measurement outcomes for a fixed number of shots per measurement~\footnote{An adaptive strategy that optimizes the number of samples in each measurement will allow to reduce the total sample count; we leave this for future work.}, the resulting sampling error---estimated from 50 independent datasets---is shown as a shaded region around the estimated mean.
As errors accumulate over time, the uncertainty in the phase increases over time and more samples are required for accurate estimation.
The accumulation of errors is particularly pronounced when the amplitude becomes small, which occurs for the product state at around $t=5$.
The sampling error is much less pronounced for the entangled initial state, owing to its large overlap with the ground state and correspondingly large value of the amplitude.

\begin{figure}[t]
    \centering
    \includegraphics[width=1.0\linewidth]{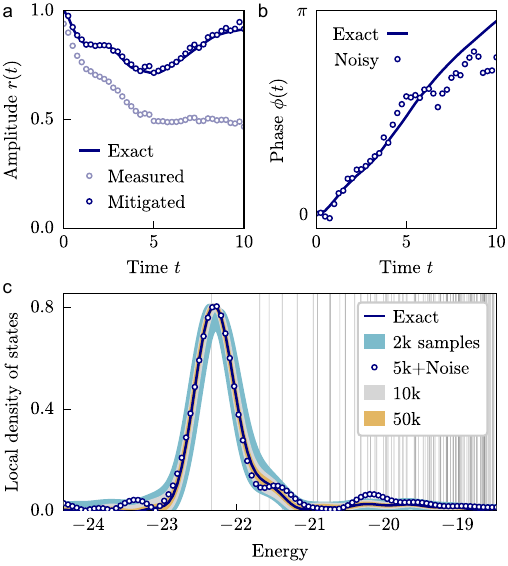}
    \caption{Application of the complex Loschmidt echo to computing the broadened local density of states (LDOS) in the presence of noise.
        We perform circuit simulations with 5,000 samples per measurement, including 
        depolarizing noise of strength corresponding to current hardware after every multi-qubit gate, considering the entangled initial state from Fig.~\ref{fig:Fig3}. 
        \textbf{(a)} The amplitude signal is exponentially damped, which can be effectively mitigated.
        \textbf{(b)} The error in the reconstructed phase is not significantly affected by the additional hardware noise, which we attribute to an inherent noise robustness of the phase gradient. 
        \textbf{(c)} We compute the LDOS with a Gaussian filter of width $\delta=0.25$, comparing the exact signal to noisy data. 
        Markers are reconstructed from the data in the previous two panels. 
        Shaded regions show the uncertainty when only shot noise is included. 
    }
    \label{fig:Fig4}
\end{figure}

We incorporate hardware errors by applying depolarizing noise at a noise strength of $\gamma_2\,=\,2\times 10^{-3}$ after every two-qubit gate, and for the single three-qubit gate choose $\gamma_3\,=\,6\times 10^{-3}$, as defined in Appendix~\ref{app:sec:Ising}, reflecting current device performance~\cite{acharya_quantum_2025}.  
The Loschmidt echo is highly sensitive to such noise. In fact, the amplitude decays swiftly from the exact values, as the data in Fig.~\ref{fig:Fig4}(a) illustrate.
The noise generally steers the dynamics away from the target state, causing the amplitude $r$ to decay exponentially compared to the noiseless value~\cite{yang_simulating_2023, yang_phase-sensitive_2024}.  
We mitigate this decay by rescaling the amplitude by $\exp(\Gamma t)$, where the rate $\Gamma$ is obtained by fitting an exponential function to the survival probability of a trivial forward and backward evolution (see Appendix~\ref{app:sec:Ising} for details).   

Remarkably, no such explicit error mitigation is required to estimate the phase gradient.
This can be readily seen from Eq.~\eqref{eq:dpg}, where
both the numerator and the denominator are subject to a similar exponential decay.
The effect of noise thus cancels to a large extent, although the variance of the estimation increases when the numerator and denominator become very small.
We observe the robustness of the phase gradient in the presence of noise in the numerical simulation in Fig.~\ref{fig:Fig4}(b), where the phase recovered for the depth-2 ansatz state without any explicit error mitigation aligns very well with the exact value.

Moreover, relative to the global (standard) Hadamard test, our sequential variant exhibits a significantly slower decay of the Loschmidt echo.
In the global Hadamard test, each gate is promoted to a controlled operation, which typically decomposes into several two-qubit gates—depending on the Hamiltonian and the available gate set—thereby increasing the effective noise per time step. 
By avoiding controlling every gate, the sequential Hadamard test accumulates less noise and permits probing longer times for the Loschmidt echo. 
A numerical comparison of the two protocols is provided in Appendix~\ref{app:Hadamard_compare}.

As an application, we use the time series of the Loschmidt echo $g(t)$
to compute the (broadened) local density of states (LDOS),
\begin{align} \label{eq:DOS}
    D(E) = \langle\psi|P_\delta(E)|\psi\rangle \approx \sum_{m=-\infty}^{\infty} c_m e^{iEt_m}g(t_m),
\end{align}
using a discretization of the Gaussian energy filter $P_\delta(E)=\exp[-(H-E)^2/(2\delta^2)]$ with a width $\delta = 0.25$. The definition of the coefficients $c_m$ and the discrete times $t_m$, together with details of the implementation is included in Appendix~\ref{app:sec:Ising}. 
The uncertainty in the LDOS combines uncertainties from both the amplitude and the phase.
Together, the LDOS derived from the noisy data is in good agreement with the exact results for the entangled initial state, accurately identifying the ground state energy [Fig.~\ref{fig:Fig4}(c)].
We calculate the number of samples required for computing the LDOS in the presence of hardware noise, as shown as a dotted line in Fig.~\ref{fig:Fig4}(c), by multiplying the length of the gate sequence (720) with the number of samples for both basis measurements of the auxiliary qubit (5,000 each) and we conclude that the few million samples can be obtained on a current superconducting qubit chip within minutes~\cite{acharya_quantum_2025}.
We remark that when an exponentially damped Loschmidt-echo amplitude is not mitigated, it leads to an additional effective line broadening in the LDOS, as detailed in Appendix~\ref{app:broadening}.

Beyond spectral properties such as the LDOS, access to complex Loschmidt echoes also allows to compute the expectation value of physical observables for both microcanonical and canonical ensembles~\cite{lu_algorithms_2021}.
Concretely, for an observable $A$ we can consider its microcanonical expectation at energy $E$ by
\begin{align}
    A_{\psi}(E)=\lim_{\delta\rightarrow 0}\frac{\langle \psi |P_\delta(E)AP_\delta(E)|\psi\rangle}{\langle \psi |P_\delta(E)^2|\psi\rangle},
\end{align}
provided the initial state has nonzero overlap with the eigenstate(s) at that energy.
In direct analogy to the LDOS in Eq.~\eqref{eq:DOS}, the microcanonical expectation value admits a representation in terms of (generalized) Loschmidt echoes $\langle\psi|e^{-iHt}|\psi\rangle$ and $\langle\psi|e^{iHt'}Ae^{-iHt}|\psi\rangle$.
Without loss of generality, $A$ can be written as sum of unitaries so that the latter amounts to evolving for a time $t$, applying a local unitary and then evolving backwards for a time $t'$. 
The phase information of these Loschmidt echoes may be evaluated without globally controlled operations by means of our proposed protocols.

\section{Discussion and outlook} \label{sec:discussion}

We have presented a toolbox of three methods to measure the phase of the Loschmidt echo without the need for a global Hadamard test. 
Our methods require only a single controlled gate per circuit (or a single layer of controlled gates for the ITE method), resulting in a near-optimally shallow circuit.
The reduction of the circuit depth comes at the cost of an increased sampling overhead, which we have rigorously bounded; the asymptotic scaling of the sample complexity is summarized in Table~\ref{tab:methods_entangled}.
Our methods are generally applicable to arbitrary quantum circuits and general many-body Hamiltonians~\footnote{A geometrically local Hamiltonian is preferred for a hardware-efficient approach, however, our methods also reduce the circuit depth for computing the Loschmidt echoes of models including non-local operators.} and our approach does not require a Hamiltonian symmetry or a reference state for which the complex Loschmidt echo is known.
Instead, we leverage the relation between the Loschmidt echo at successive times.
This strategy is particularly suitable for spectroscopy applications, which require a complete time series of the complex Loschmidt echo.
We further highlight that our methods can be applied to any quantum computation platform, including hybrid analog-digital devices.
Previous algorithms relying on globally controlled dynamics are not compatible with the analog Hamiltonian dynamics on quantum simulators.

On quantum computers with fixed qubit connectivity, the need for auxiliary qubits may necessitate a number of swap operations or may restrict the connectivity of Hamiltonians that can be efficiently implemented in the near term.
Platforms with a more flexible connectivity, such as trapped ions or neutral atom arrays, largely avoid this limitation~\cite{bruzewicz_trapped-ion_2019, monroe_programmable_2021, henriet_quantum_2020, kaufman_quantum_2021}.
Our methods therefore yield the largest resource savings when qubit connectivity is limited.
In the global Hadamard test---where a single qubit coherently controls operations at multiple, spatially separated sites---the necessary swap operations introduce a multiplicative depth overhead.
Typically, each controlled gate must also be decomposed into the native gate set, contributing a further factor to the depth.
By turning these overheads into additive terms, our methods can significantly reduce the depth.
Since, prior to fully fault-tolerant quantum computation, hardware imperfections impose a practical ceiling on the executable depth~\cite{aharonov_limitations_1996}, this depth reduction enables probing complex Loschmidt echoes up to larger evolution times---which translates into more precise phase estimation.

As with all quantum phase estimation techniques, our approach requires an initial state with sufficient overlap with the eigenstates of interest---a condition that typically necessitates entanglement. 
Hence, a key strength of our methods is their applicability to arbitrary quantum states, rather than being limited to product states.
This is particularly relevant for local Hamiltonians, for which the Loschmidt echo amplitude of generic product states decays to an exponentially small value in a short time $t\sim1/\sqrt{n}$~\cite{hartmann_gaussian_2004, kuwahara_connecting_2016, anshu_concentration_2016}.
In contrast, states with a significant overlap with an eigenstate can maintain a robust Loschmidt echo amplitude, rendering phase integration more efficient.
Our method is compatible with all reversible state preparation methods, in particular with Trotterized quantum adiabatic algorithms~\cite{farhi_quantum_2000}, variational quantum algorithms~\cite{cerezo_variational_2021}, or with circuits preparing a Hartree-Fock ansatz, a widely used choice for the initial state for quantum chemistry applications~\cite{google_ai_quantum_and_collaborators_hartree-fock_2020}.
If the state preparation cannot be uncomputed, our methods can be adapted into a two-copy protocol where the state projection is realized with a destructive SWAP-test, requiring only local Bell state measurements (see~Appendix~\ref{app:project}).

While our numerical case study has focused on the conceptually simplest case of estimating energy eigenvalues, access to the complex Loschmidt echo also opens up an avenue to compute observables in thermal equilibrium for any Hamiltonian, including those afflicted by the sign problem in quantum Monte Carlo methods~\cite{poulin_sampling_2009, lu_algorithms_2021, huggins_unbiasing_2022}.
This capability is relevant for a broad range of applications, including studies of topological order~\cite{semeghini_probing_2021, iqbal_non-abelian_2024} and strongly correlated models such as the Fermi-Hubbard model~\cite{chalopin_probing_2024, evered_probing_2025, xu_neutral-atom_2025}.
Current quantum computers have recently demonstrated the ability of carrying out quantum simulation tasks at or beyond the boundary of classical computers~\cite{kim_evidence_2023, king_beyond-classical_2025, andersen_thermalization_2025, haghshenas_digital_2025, abanin_constructive_2025}. 
Since our methods require few additional quantum resources, they are immediately applicable to these settings. 
By recovering the phase of the Loschmidt echo, which is an inherently nonlocal quantity, our methods may, in fact, push quantum computers further beyond the reach of classical computation. 
The nonlocality enters our protocols through the projective measurement onto the initial state, which renders classical simulation tailored for local observables impractical~\cite{rudolph_classical_2023,wild_classical_2023,shao_simulating_2024, tindall_efficient_2024, begusic_real-time_2025}.
Our work thus creates an avenue towards performing quantum phase estimation in classically intractable regimes.

\begin{acknowledgments}
We thank Tom O'Brien, Johannes Feldmeier, Nazlı Uğur Köylüoğlu, Sirui Lu, Yorgos Styliaris, Rahul Trivedi and Yilun Yang for insightful discussions. 
B.F.S.~and J.I.C.~acknowledge funding from the Federal Ministry of Education and Research Germany (BMBF) via the project FermiQP (13N15889). 
Work at MPQ is part of the Munich Quantum Valley, which is supported by the Bavarian state government with funds from the Hightech Agenda Bayern Plus. 
Work at Harvard was supported by the US Department of Energy (DOE Gauge-Gravity, grant number DE-SC0021013, and DOE Quantum Systems Accelerator, grant number DE-AC02-05CH11231), the National Science Foundation (grant number PHY-2012023), the Center for Ultracold Atoms (an NSF Physics Frontiers Center), and the Army Research Office MURI (grant number W911NF2010082) and Wellcome Leap Foundation Quantum for Bio program. 
\end{acknowledgments}

%\clearpage
\appendix

\section{Implementing the state projection} \label{app:project}

We comment on the realization of projecting onto the initial state, which is required for all three protocols presented in the main text.
Quantum devices are generally initialized in an unentangled, computational basis state, e.g.~$\ket{\text{init}}=\ket{0}\otimes\ket{0}\otimes\dots\otimes\ket{0}$.
After initialization, the probe state for quantum phase estimation $\ket{\psi}$ needs to be realized by a state-preparation method of choice, e.g.~a variational quantum algorithm~\cite{cerezo_variational_2021}, or adiabatic state preparation~\cite{lidar_adiabatic_2009, schiffer_adiabatic_2022}. 
The action of these methods can then be described by a unitary transformation $\ket{\psi} = U \ket{\text{init}}$.
Hence, the projection onto the initial state $\ket{\psi}$ can be realized by the reverse transformation $U^\dag$ before a computational basis measurement:
\begin{align}
    |\psi\rangle\langle\psi| = U|\text{init}\rangle\langle\text{init}|U^\dag.
\end{align}
The computational basis projection is performed by simply taking snapshots in the computational basis and counting the occurrences of the all-zero state $\ket{\text{init}}$.

For some scenarios, the state preparation cannot be reversed. 
One such case is non-unitary state preparation, using dissipation~\cite{mi_stable_2024,molpeceres_quantum_2025} or measurements~\cite{piroli_approximating_2024}.
Also, when implementing a not purely adiabatic sweep on an analog simulator, the inverse operation $U^\dag$ is often not available.
In this case, a two-copy protocol can be used implementing a destructive SWAP test using Bell pair measurements on pairs of qubits~\cite{huggins_virtual_2021}. Mathematically, this projection allows to compute $\text{Tr}[\sigma\rho]$ with $\sigma$ and $\rho$ the density matrices of two quantum states.

\section{Details of the error analysis}

In the derivation of the sample complexity for the three methods described in the main text, some details have been omitted for brevity of the presentation. 
We include them here.

\subsection{Error of the sequential Hadamard test} \label{app:ssec:sequ}
We include details on the derivation of the sampling error for the sequential Hadamard test.
In each iteration of the protocol, we consider the circuit in Fig.~\ref{fig:sequ_grad_1}.
For the error analysis, we define the observables $A_\alpha = \sigma^\alpha \otimes |\psi\rangle\langle\psi|$, with $\alpha \in \{x,y\}$.
The first and second moments of $A_\alpha$ are given as
\begin{align}
    \langle A_x \rangle_l &= \langle \sigma^x \otimes |\psi\rangle\langle\psi| \rangle_l = \Re[g_l g_{l-1}^*],\\
    \langle A_y \rangle_l &= \langle \sigma^y \otimes |\psi\rangle\langle\psi| \rangle_l =\Im[g_l g_{l-1}^*], \\
    \langle A_\alpha^2\rangle_l &= \langle \mathbb{I}\otimes |\psi\rangle\langle\psi| \rangle_l =(r_l^2 + r_{l-1}^2)/2.
\end{align}
We compute the variance of the phase difference
\begin{align}
    \delta \phi_l = \tan^{-1}[\Im(g_l^{}g_{l-1}^*)/\Re(g_l^{}g_{l-1}^*)].
\end{align}
First, we define the shorthands $x_l = \langle A_x\rangle_l$, and $y_l = \langle A_y\rangle_l$ and note that $x_l^2+y_l^2=r_l^2 r_{l-1}^2$.
The variance of the phase difference $\delta\phi_l$ is
    \begin{align} \label{app:eq:phasediff}
    (\Delta[\delta\phi_l])^2 
    &= \frac{
      (\partial_{x}\phi_l)^2(\Delta A_x)^2
    + (\partial_y\phi_l)^2(\Delta A_y)^2}{M_l},
\end{align}
where the variance of the observables is given by
\begin{align}
    (\Delta A_\alpha)^2
    &= \langle A_\alpha^2\rangle_l - \langle A_\alpha\rangle_l^2 \nonumber\\
    &= (r_l^2 + r_{l-1}^2)/2 - \langle A_\alpha\rangle_l^2.
\end{align}
We recall that $\frac{\di}{\di x}\tan^{-1}(x)=(1+x^2)^{-1}$ and state the partial derivatives
\begin{align}
    \partial_{x}\phi_l&=\frac{\partial\phi_l}{\partial x_l}
    = -\frac{y_l}{x_l^2 + y_l^2} = -\frac{y_l}{r_l^2 r_{l-1}^2}, \\
    \partial_{y}\phi_l&=\frac{\partial\phi_l}{\partial y_l}
    = \frac{x_l}{x_l^2 + y_l^2}=\frac{x_l}{r_l^2 r_{l-1}^2}.
\end{align}
Inserting the derivatives into the expression for the error in Eq.~\eqref{app:eq:phasediff}, yields
\begin{align}
    (\delta\phi_l)^2
    &= \frac{y_l^2(r_l^2+r_{l-1}^2-2x_l^2)
    + x_l^2(r_l^2+r_{l-1}^2-2y_l^2)}{2M_l^{}r_l^4 r_{l-1}^4}
   \nonumber\\
   &\leq \frac{r_l^2r_{l-1}^2(r_l^2+r_{l-1}^2)}{2M_l^{}r_l^4 r_{l-1}^4}
   \nonumber\\
    &= \frac{r_l^2 + r_{l-1}^2}{2M_l^{}r_l^2 r_{l-1}^2}.
\end{align}
Hence, we arrive at the bound
\begin{align} \label{app:eq:delta_phi}
    (\Delta[\delta\phi_l])^2 \leq \frac{1}{2M_l}\left(\frac{1}{r_l^2} + \frac{1}{r_{l-1}^2} \right).
\end{align}

Alternatively to considering the phase separately from the amplitude, we could also proceed by directly reconstructing the complex Loschmidt echo, as prescribed by Eq.~\eqref{eq:recursive}. 
Then, instead of adding and subtracting the phase difference $\delta \phi_l$, one multiplies and divides the quantity $g_l g_{l-1}^*$.
Hence, to analyze the propagation of the sampling error, we consider the multiplicative error of $g_l g_{l-1}^*$.
The squared error $(\Delta [g_l g_{l-1}^*])^2$ follows as the sum of the variance of $A_x$ and $A_y$:
\begin{align}
    \Delta [g_l g_{l-1}^*]^2 &= \frac{1}{M_l}\left(\langle A_x^2\rangle_l + \langle A_y^2\rangle_l - \langle A_x\rangle^2_l - \langle A_y\rangle^2_l\right) \nonumber\\
    &= \frac{1}{M_l}\left(r_l^2 + r_{l-1}^2 - r_l^2 r_{l-1}^2\right).
\end{align}
Then, for the squared multiplicative error of $g_l g_{l-1}^*$ we obtain
\begin{align}
    \frac{ (\Delta [g_l g_{l-1}^*])^2}{ (r_l r_{l-1})^2} &= \frac{r_l^2 + r_{l-1}^2 - r_l^2 r_{l-1}^2}{M_l^{}r_l^2 r_{l-1}^2} \nonumber\\
   &= \frac{1}{M_l}\left(\frac{1}{r_l^2} + \frac{1}{r_{l-1}^2} - 1\right) \nonumber\\
    &\leq \frac{1}{M_l}\left(\frac{1}{r_l^2} + \frac{1}{r_{l-1}^2} \right).
\end{align}
Hence, both approaches yield a similar scaling of the error. For the rest of the derivation we refer to the main text.

\subsection{Error of the direct phase gradient method} \label{app:ssec:dpg}

The output of the quantum circuit in Fig.~\ref{fig:directGradCircuit} yields the quantity
\begin{align}
    a_{j,l} = \frac{1}{2}\bra{\psi} \exp(iHt_l) \{P_j, |\psi\rangle\langle\psi|\} \exp(-iHt_l) \ket{\psi},
\end{align}
at a time $t_l$, where $l \in \{1, 2, \ldots, N_\text{times} \}$. 
The Hamiltonian is $H=\sum_j \lambda_j P_j$, with $\lambda_j$ a scalar and $P_j$ a Pauli string.
For the error analysis we write the probability $p_l= r^2_l = |\langle\psi|\exp(-i Ht_l)|\psi\rangle|^2$ for a successful projection onto the initial state. 
The phase gradient is then given as
\begin{align}
   \phi'_l= \left. \frac{\di \phi(t)}{\di t} \right|_{t = t_l}  = -\frac{\sum_j \lambda_j a_{j,l}}{p_l}
\end{align}
and error in the phase gradient is:
\begin{align}
    (\Delta \phi'_l) &\approx  (\phi'_l)^2 \left[\left(\frac{\lambda_j\Delta(\sum_j a_{j,l})}{\sum_j a_{j,l}} \right)^2+ \left(\frac{\Delta p_l}{p_l}\right)^2 \right] \nonumber\\
    &=\frac{(\sum_j a_{j,l})^2}{p_l^2} \left[ \frac{(\sum_j \lambda_j\Delta a_{j,l})^2}{(\sum_j a_{j,l})^2} +\frac{(\Delta p_l)^2}{p_l^2}  \right] \nonumber\\
    &= \frac{(\sum_j \lambda_j \Delta a_{j,l})^2}{p_l^2} +  (\phi'_l)^2 \frac{(\Delta p_l)^2}{p_l^2}\nonumber\\
    &\approx\frac{1}{p_l^2} \sum_{j=1}^K \lambda_j^2 (\Delta a_{j,l})^2 + \left( \phi'_l \right)^2 \frac{(\Delta p_l)^2}{p_l^2}, 
\end{align}
where we assumed that all errors are sufficiently small for the error propagation formula to apply~\cite{taylor_introduction_1997}.

To compute the error term $(\sum_i \Delta a_{j,l})^2$, we consider the first two moments of the observable $A_x$:
\begin{align}
    \langle A_{x}\rangle_{j,l}^2 &=\langle \sigma^x\otimes |\psi\rangle\langle\psi| \rangle_{j,l}^2 = a_{j,l}^2,\\
    \langle A_x^2\rangle_j &= \frac{1}{2}\langle \mathbb{I}\otimes |\psi\rangle\langle\psi| \rangle_{j,l} \nonumber\\
    &= \frac{1}{2}\left(p_l + |\langle\psi|P_j \exp(-it_lH)|\psi\rangle|^2\right).
\end{align}

If we run $M^{(a)}_{j,l}$ repetitions of the circuit in Fig.~\ref{fig:directGradCircuit}, the variance of $a_{j,l}$ is given by
\begin{align}
    (\Delta a_{j,l})^2 &=  \frac{\langle A_x^2\rangle_{j,l} - \langle A_{x}\rangle_{j,l}^2 }{M^{(a)}_{j,l}}\nonumber\\
    &= \frac{1}{2 M^{(a)}_{j,l}} \left( p_l + \left| \braket{\psi | P_j e^{-i H t_l} | \psi} \right|^2 - 2 a_{j,l}^2 \right).
\end{align}
To measure $p_l$, we remove the auxiliary qubit and the controlled Pauli operation from the circuit. The probability that the projection onto $\ket{\psi}$ succeeds is equal to $p_l$ and $\Delta p_l$ follows from the variance of the corresponding binomial distribution. With $M^{(r)}_l$ measurements, it is given by $(\Delta p_l)^2 = p_l \left( 1 - p_l \right)/M^{(r)}_{l}$.

To arrive at more transparent expressions, we bound these variances by $(\Delta a_{j,l})^2 \leq 1/M^{(a)}_{j,l}$ and $(\Delta p_l)^2 \leq p_l / M^{(r)}_l$. We further note that $|\phi_l'| = \bigO{n}$, assuming that the Loschmidt echo is nonvanishing~\cite{yang_phase-sensitive_2024}. Hence, the measurement numbers $M^{(r)}_l = \bigO{n^2 / (\eta^2 p_l)}$ and $M^{(a)}_{j,l} = \bigO{n / (\eta^2 p_l^2)}$ are sufficient to compute the phase gradient $\Delta \phi_l'$ with additive error $\eta$. The total number of measurements for a single time step then satisfies $M_l = \sum_{j = 1}^K M^{(a)}_{j,l} + M^{(r)}_l = \bigO{(n / \eta p_l)^2}$.

In addition to the statistical uncertainty in the phase gradient, the estimation of the final phase incurs a numerical integration error. Denoting the integration error by $\Delta \phi_I$, the total error satisfies $\Delta \phi \leq N_\text{times}^{1/2} \eta + \Delta \phi_I$, where we chose the same statistical error $\eta$ at each time step and used the fact that the statistical errors are uncorrelated.
For the Newton-Cotes formula~\footnote{In practice, other numerical integration methods with uneven spacing, such as those based on Gaussian quadrature rules, could yield better integration results, but bounding the integration error is more challenging.}, the integration error is $\Delta\phi_I = \bigO{n t  (t/N_\text{times})^{2 s}}$, where $s$ is an integer that depends on the order of the formula ($s = 1$ for the trapezoidal rule, $s = 2$ for Simpson's rule)~\cite{quarteroni_numerical_2007}. Hence, $N_\text{times} = \bigO{(n t / \epsilon)^{1/2s} \, t}$ discrete time points and $M_l = \bigO{(n t / \epsilon)^{2 + 1/2s} t^{-1} p_l^{-2}}$ respective measurements are sufficient to ensure that $\Delta \phi \leq \epsilon$. The total number of measurements is thus bounded by
\begin{equation}
    M_\mathrm{tot} = \sum_{l = 1}^{N_\text{times}} M_l = \bigO{(n t / \epsilon)^{2 + 1/s} r_\text{min}^{-4}}.
\end{equation}

\subsection{Error of the ITE phase gradient method} \label{app:ssec:ite}

We start the discussion of the error in the ITE phase gradient method from the circuit in Fig.~\ref{fig:LocalBE}.
After postselecting on the auxiliary qubit being in the $\ket{0}$ state, the circuit prepares the (unnormalized) state
\begin{align} \label{eq:ITE_cos_app}
    [\cos^2(\theta/2)\mathbb{I} \pm \sin^2(\theta/2)P_j] \ket{\psi}.
\end{align}
By comparing this expression with the identity
\begin{equation}
    e^{\pm \tau \lambda_j P_j} = \cosh(\tau \lambda_j) \mathbb{I} \pm \sinh(\tau \lambda_j) P_j,
\end{equation}
we can see that the postselected circuit thus implements the non-unitary operation $e^{- \tau |\lambda_j|} e^{\pm \tau \lambda_j P_j}$ if we choose $\theta$ such that $\cos^2 (\theta / 2) =  e^{-\tau |\lambda_j|} \cosh(\tau \lambda_j)$ and $\sin^2 (\theta / 2) =  e^{-\tau |\lambda_j|} \sinh(\tau |\lambda_j|)$. By applying a similar circuit to each term in the Hamiltonian, we are thus able to apply the operation $\prod_j e^{- \tau |\lambda_j|} e^{\pm \tau \lambda_j P_j} \approx e^{- \tau \sum_j |\lambda_j|} e^{\pm H \tau}$.

To determine the sample complexity, we first analyze the postselection overhead from the local block-encodings. 
The success probability of each measurement is given by the norm squared of the state in Eq.~\eqref{eq:ITE_cos_app}. For small values of $\tau$, the probability that all measurements succeed is thus given by
\begin{align}
    &P_\mathrm{succ} = \prod_j e^{- 2 |\lambda_j| \tau} \braket{\psi | e^{\pm 2 \tau \lambda_j P_j} | \psi} \nonumber\\
    &= 1 - 2 \tau \sum_j \left( |\lambda_j| \mp \lambda_j \braket{\psi | P_j | \psi} \right) + \bigO{\tau^2}.
\end{align}
As there are $\bigO{n}$ Hamiltonian terms, the failure probability for a short ITE is $P_\text{fail} = \bigO{n\tau}$. If any one of the auxiliary qubits heralds failure, the protocol is restarted.
To achieve a constant success probability, independent of system size, we require that the imaginary-time duration decreases with system size, $\tau = \bigO{1/n}$. 
This is consistent with the requirement that $\tau$ be small to limit the Trotter error, which we discuss in more detail below.

As for the previous method, shot noise leads to a statistical uncertainty in the phase gradient $\phi'_l = \left. \di \phi(t) / \di t \right|_{t = t_l}$. With the shorthand $p_{l}^\pm = r(t_l\pm i\tau)^2$, the expected variance of $\phi'_l$ is given by
\begin{align}
    (\Delta \phi'_l)^2 \approx \frac{1}{(4 \tau)^2} \left[ \left(\frac{\Delta p_l^+}{p_l^+} \right)^2 + \left( \frac{\Delta p_l^-}{p_l^-} \right)^2 \right] ,
\end{align}
assuming the uncertainties $\Delta p_l^\pm$ are sufficiently small~\cite{taylor_introduction_1997}.
Using the above implementation of the imaginary time evolution, it is possible to measure $q_l^\pm = e^{- 2\tau \sum_j |\lambda_j|} p_l^\pm$ by performing a projective measurement onto $\ket{\psi}$. Since the measurement outcomes follow a binomial distribution, the corresponding uncertainty satisfies $(\Delta q_l^\pm)^2 \leq q_l^\pm / M_l^\pm$, where $M_l^\pm$ denotes the respective number of measurements. It follows that $(\Delta p_l^\pm)^2 \leq e^{2 \tau \sum_j |\lambda_j|} p_l^\pm / M_l^\pm$. The exponential prefactor can be ignored as $\tau = \bigO{1/n}$. Hence, $M_{l}^\pm=\bigO{1/(\tau^2\eta^2 p_{l}^\pm)}$ samples are sufficient to ensure that the error $\Delta \phi'_l$ due to shot noise is bounded from above by $\eta$.

As in section~\ref{ssec:DirectGrad}, we numerically integrate the phase gradient using a Newton-Cotes formula with $N_\text{times} = \bigO{(n t / \epsilon)^{1/2s} \, t}$, such that the integration error is less than~$\epsilon$. 
Accordingly, we choose $\eta = \bigO{\epsilon / N_\text{times}^{1/2}}$ to bound the accumulated statistical uncertainty by~$\epsilon$. 
In addition to the statistical uncertainty and numerical integration error, the phase gradient incurs a systematic error due to the first-order Trotter decomposition of the imaginary time evolution and the finite-difference approximation of the derivative in Eq.~\eqref{eq:ite}. 
Assuming that the Loschmidt echo and a closely related quantity do not vanish, both errors can be shown to be of size $\bigO{n \tau^2}$~\cite{yang_phase-sensitive_2024}, which may accumulate during the numerical integration to $\bigO{N_\text{times} n \tau^2}$. By choosing $\tau = \bigO{(\epsilon / t)^{(1 + 1/2s)/2} / n}$, we ensure that this total error is bounded by $\epsilon$ while satisfying the above requirement $\tau = \bigO{1/n}$. Combining these results yields the total number of samples
\begin{align}
    M_\text{tot} = \sum_{l=1}^{N_\text{times}} (M_l^+ + M_l^-)=\bigO{\frac{N_{\text{times}}^2}{\tau^2\,\epsilon^2\,r_{\pm,\min}^2}},
\end{align}
where $r_{\pm, \text{min}} = \min_{t' \in [0, t]} \{ r(t' \pm i \tau) \}$.
By plugging in the bound for $\tau$, we finally obtain
\begin{equation}
    M_\text{tot} = \bigO{n^{2 + 1/s} (t / \epsilon)^{3 + 3/2s} r_{\pm, \text{min}}^{-2}} .
\end{equation}

\section{Optimized sequential Hadamard test} \label{app:cutting}

We include a discussion of the optimized sequence for the sequential Hadamard method. 
In the main text, we mention that there is a freedom how to choose the sequence of gates in the sequential Hadamard test to compute the Loschmidt echo. 
An optimized sequence allows to reduce the total number of samples and can also reduce the circuit depth.
The key observation is that in the sequential control technique, starting from the full circuit with a single controlled gate, we effectively remove the controlled gate from the circuit for the next iteration and control another gate. This process is repeated until there are no gates left in the circuit. 

Let us consider a one-dimensional brick-wall circuit that approximates the dynamics of a geometrically 2-local Hamiltonian $H=H_A+H_B$. 
The Hamiltonian terms of $H_A=\sum_{j\,\text{odd}} h_{j,j+1}$  act on the sites $(1,2)$, $(3,4)$, etc., while those of $H_B=\sum_{j\,\text{even}} h_{j,j+1}$ act on $(2,3)$, $(4,5)$, etc. 
We consider a second-order Trotter decomposition to approximate the dynamics.
The circuit is composed of alternating layers implementing the gates corresponding to $H_A$ and $H_B$, respectively.
When a gate is removed from the circuit, we can often merge neighboring gates by multiplying the two unitaries.
This reduces the total number of gates and may also lower the number of entangling gates when considering a decomposition into a native gate set.
Each layer comprises two moments, in which all odd or all even gates are applied simultaneously. 
Let the circuit depth $D$ be the number of layers.
Our optimization proceeds by removing, one gate at a time, all gates from every second moment until all moments corresponding to $H_A$ are eliminated.
At that point only the $H_B$ gates remain---which is effectively half of the total gates -- and they can be fully merged into a single moment.
Whereas the original sequence required on the order of $ND$ circuits, the optimized sequence reduces this number to about $ND/2$ circuits.

We note that if we seek to compute the time series of the Loschmidt echo, e.g.~for spectroscopy applications, then the sequence should be optimized so that its intermediate circuits correspond to proper evolution times.
At some particular time $t=3\nu$, this evolution may be implemented by the circuit
\begin{align}
    e^{-iHt} \approx U_{A}^{1/2} U_{B}^{} U_{A}^{} U_{B} U_{A}^{} U_{B}^{} U_{A}^{1/2},
\end{align}
where $U_A=\exp(-iH_A\nu)$ and $U_B=\exp(-iH_B\nu)$, with a small Trotter step $\nu$.
Each of these layers is composed of local two-qubit gates in our case.
A simple strategy that preserves proper times while reducing the gate count is to start from the full circuit at the largest time $t$ and remove a full Trotter layer $U_AU_B$ in the middle of the circuit: first, we remove a central $U_B$ moment, which yields the circuit
\begin{align}
    U_{A}^{1/2}U_{B}^{} U_{A}^2 U_{B}^{} U_{A}^{1/2}.
\end{align}
We thereby merge gates from the previous or subsequent layer, wherever possible. 
Next, we remove the adjacent $U_A$ moment, such that we are left with the circuit that realizes the dynamics for an earlier time step $t'=t-\nu$:
\begin{align}
    e^{-iHt'} \approx U_{A}^{1/2} U_{B}^{} U_{A}^{} U_{B}^{} U_{A}^{1/2}.
\end{align}
This is the approach we use for the numerical case study of the Ising model in Sec.~\ref{sec:numerics}.

The above method effectively cuts the circuit along the spatial direction.
Another approach is to optimize the gate sequence by slicing along the time direction.
This can also yield significant savings in the sample cost if the initial state is a product states or prepared by a shallow circuit, as detailed below in App.~\ref{app:product_sht}.

\section{Additional remarks for initial product states}

When the initial state $\ket{\psi}$ is a product state, substantial simplifications of our protocol are possible.
For some methods, the simplifications also apply for special classes of weakly entangled states.

\subsection{Sequential Hadamard test} \label{app:product_sht}

The sequential Hadamard method allows to optimize the order of the controlled gates, which we exploit here to reduce the sample complexity when the initial state is a product state.
If we remove all gates of the circuit that entangle two subsystems, we can effectively cut the circuit in half, by cutting along the time dimension. 
Repeatedly cutting the remaining system, we are eventually left with computing the Loschmidt echo of a few-qubit system, which can be done on a classical computer.

By assuming \emph{translational invariance}, the remaining subsystems are identical, allowing to save significant resources.
For one-dimensional systems the length of the cut, i.e.~the number of gates that we need to remove to perform a single cut, is $\bigO{D}$ for a circuit depth $D$.
Hence, it is independent of the system size.
Using bi-sections, the number of cuts that are required until the remaining circuit can be classically simulated is a logarithmic function of the system size.  
Hence, in one dimension, we have $\bigO{D\log[n]}$ many circuits that we need to execute to compute the Loschmidt echo up to time $t$.

In more than one dimensions, the length of the cut generally depends on the system size. 
For a translationally-invariant Hamiltonian on a $d$-dimensional regular lattice, with a side length of $n^{1/d}$, a hyperplane cut has a boundary $\bigO{n^{(d-1)/d}}$.
The number of circuits in the sequential Hadamard test is therefore $\bigO{tn^{(d-1)/d}}$.
Then, the total sample complexity is
\begin{align}
     M_{\text{tot, product}} &= \mathcal{\widetilde{O}} \left((n^{(d-1)/d} D/\epsilon)^{2}  r_\text{min}^{-2}\right),
\end{align}
where the $\mathcal{\widetilde{O}}$-notation includes the logarithmic dependence on the system size.

We highlight that this technique can also be applied if the initial state is prepared by a shallow circuit $V$, such that $V\ket{\text{init}}=\ket{\psi}$.
Instead of $\langle \psi|U|\psi\rangle$, one simply also computes the Loschmidt echo for the transformed unitary $\langle \text{init}|V^\dag UV|\text{init}\rangle$, slightly increasing the length of each cut.

\subsection{Direct phase gradient method} \label{app:product_dpg}

We explain in the main text that, when the initial state is a product state, the direct phase gradient method can be used without auxiliary qubits. 
As an alternative to preparing both states $\ket{\varphi_j^\pm} = (\ket{\psi} \pm P_j \ket{\psi})/\sqrt{2}$, only one of them can be used.
We can see how the controlization of the gate can be avoided by considering the superposition
\begin{align}
    \ket{\varphi_j^+} = \frac{1}{\sqrt{2}}(\ket{\psi} + h_j \ket{\psi}).
\end{align}
This state can be easily prepared by a local gate.
Note that we may directly consider $h_j$, instead of $P_j$.
We can measure the probability
\begin{align}
     |\langle \varphi_j | \psi(t_l)\rangle|^2 
    =& \frac{1}{2}|\langle \psi | \psi(t_l)\rangle|^2 + \frac{1}{2}|\langle \psi | h_j | \psi(t)\rangle|^2 \nonumber\\
    &+ \frac{1}{2}\langle\psi(t_l)|\{h_j,|\psi\rangle\langle\psi|\}|\psi(t_l)\rangle,
\end{align}
writing $\ket{\psi(t_l)}=\exp(-iHt_l)\ket{\psi}$.
Hence, this approach also allows to evaluate the phase gradient
\begin{align}
    \left. \frac{\di \phi(t)}{\di t} \right|_{t = t_l} =& \frac{1}{2 r_{l}^2} \sum_j \big( 2|\langle\varphi_j|\psi(t_l)\rangle|^2  -|\langle{\psi}|h_j|\psi(t_l)\rangle|^2  \big) 
     - \frac{1}{2}
\end{align}
 by only measuring the survival probabilities without controlled operations.

\subsection{ITE phase gradient method} \label{app:product_ite}

For product states, the ITE phase gradient method as proposed in~\cite{yang_phase-sensitive_2024} can be used.
For the sample complexity to obtain the Loschmidt echo $g(t)$ within additive error $\epsilon$ we obtain
\begin{align}
    M_\text{tot, product} = \bigO{n^{1 + 1/s} (t / \epsilon)^{3 + 3/2s} r_{\pm, \text{min}}^{-2}},
\end{align}
with $s$ the integer depending on the order of the numerical integration formula ($s = 1$ for the trapezoidal rule, $s = 2$ for Simpson's rule)~\cite{quarteroni_numerical_2007}.
This result differs slightly from Ref.~\cite{yang_phase-sensitive_2024}:
besides including the integration error, we obtain a lower sample complexity in the time $t$, because we use that the statistical noise is uncorrelated.

\section{Reference-state method without global superpositions}\label{app:ref_state}

Our methods do not require a reference state with a known complex Loschmidt echo. However, our approach can also be beneficial when such a reference state is available. In particular, instead of tracking the phase of the initial state during the evolution, we can apply our approach to track the change of the phase of the Loschmidt echo during the preparation of the initial state from a given reference state. Unlike conventional reference-state methods, this does not require the preparation of a superposition of the initial state and the reference state.

Concretely, consider a state $ |\psi\rangle$ prepared from a state $|0^n\rangle$ with $M$ gates as
\begin{align}
  |\psi\rangle = W|0^n\rangle,\qquad
  W = w_M\cdots w_2 w_1.
\end{align}
Each gate $w_L$ is assumed to act locally on neighboring qubits. 
Here, $|0^n\rangle$ acts as a reference state, where the phase of the Loschmidt echo is known.
We define the notation for the intermediate state
\begin{align}
  |\psi_L\rangle =w_L\cdots w_1|0^n\rangle,
\end{align}
such that $|\psi_M\rangle = |\psi\rangle$. 
Our goal is to compute the Loschmidt amplitude
\begin{align}
  g_M = \langle \psi_M|U|\psi_M\rangle = r_M e^{i\phi_M}.
\end{align}
We may then relate $g_{L}=\langle \psi_{L}|U|\psi_{L}\rangle$ to $g_{L-1} = \langle \psi_{L-1}|U|\psi_{L-1}\rangle$ using the sequential Hadamard test.
We proceed by first considering a circuit where we control the gate $w_{L}$ with an auxiliary qubit to obtain $g_{L}\langle \psi_{L}|U|\psi_{L-1}\rangle^*$. 
As a second step, we control the gate $w_L^\dag$ and compute $\langle \psi_{L}|U|\psi_{L-1}\rangle g_{L-1}^*$.
Both steps can be combined such that the two gates are controlled in one circuit, which yields the quantity $g_{L}^{} g_{L-1}^*$. 
By repeating the procedure for all $M$ gates $\{w_L\}_L$, we obtain the full sequence $g_{M}^{} g_{M-1}^*, \dots, g_{1}^{} g_{0}^*$. 
Here, the Loschmidt echo of the reference state $g_{0}=\langle 0^n|U|0^n\rangle$ is known by assumption.

While the sample complexity depends on the amplitudes of the intermediate states $\{|g_L|\}_L$, the reference state approach might be particularly useful when the number of gates $M$ for preparing the initial state is much smaller than the number of gates in the decomposition of the unitary~$U$.

\hfill

\section{Control-free variant for phase-sensitive measurements} \label{app:no_ancilla}

We describe a protocol without controlled operations or auxiliary qubits for computing the complex Loschmidt echo $\langle\psi|P_jU|\psi\rangle$ for a Pauli operator $P_j$ and a unitary $U$ when $\langle\psi|U|\psi\rangle$ is known.
The Loschmidt echo then directly provides the coefficients $a_j$ for the direct phase gradient method.
Similarly, knowledge of $\langle\psi|PU|\psi\rangle$ allows us to realize the sequential Hadamard test without an auxiliary qubit: 
instead of controlling the gate $u_L=e^{i\alpha P}$, we obtain $\langle\psi|u_LU|\psi\rangle$ via the relation
\begin{align}
    \langle \psi|e^{i\alpha P}U|\psi\rangle = \cos(\alpha)\langle\psi|U|\psi\rangle \!+\! i\sin(\alpha)\langle\psi|PU|\psi\rangle. 
\end{align}
We restrict the analysis to multi-qubit Pauli rotation gates, as it is possible to expand any Hermitian operator in the Pauli basis.

We assume that the non-zero Loschmidt echo $\langle\psi|U|\psi\rangle$ is known, and we seek to obtain $\langle\psi|PU|\psi\rangle$.
We consider the two expressions
\begin{align}
    x(\theta) &= \langle \psi|e^{i\theta P}U|\psi\rangle, \label{eq:x}\\
    y(\theta_1, \theta_2) &= \langle \psi|e^{i\theta_1 P_1}e^{i\theta_2 P_2}U|\psi\rangle. \label{eq:y}
\end{align}
The exponentials are Pauli rotations with the respective Pauli strings $P$, $P_1$ and $P_2$. 
One may choose different angles for the two rotations in the expression for $y$, however, a single angle will suffice for our purposes.
Crucially, the Pauli strings are chosen such that they factor as
\begin{align}
    P_1P_2=P.
\end{align}
The Loschmidt echo $r=\langle\psi|U|\psi\rangle\neq0$ may be chosen real, as we can arbitrarily redefine the global phase. 
We further use the notation 
\begin{align}
    p&=\langle\psi|PU|\psi\rangle, \\
    p_1&=\langle\psi|P_1U|\psi\rangle, \\
    p_2&=\langle\psi|P_2U|\psi\rangle
\end{align} 
for the other Loschmidt echoes in the calculation.
In our protocol, we allow for measurements of the parametrized amplitudes $|x(\theta)|^2$ and $|y(\theta)|^2$. 
Expanding Eq.~\eqref{eq:x}, we define the shorthands $c = \cos(\theta)$ and $s = \sin(\theta)$ and obtain
\begin{align}
    |x(\theta)|^2 &= c^2r^2-2csr\,\Im[p] + s^2|p|^2,\label{eq:x_sq}
\end{align}
From Eq.~\eqref{eq:x_sq}, by measuring $|x(\tfrac{\pi}{4})|^2$ and $|p|^2=|x(\tfrac{\pi}{2})|^2$, we can immediately obtain 
\begin{align}
    \Im[p] = \frac{1}{2r}(r^2+|p|^2 - 2|x(\tfrac{\pi}{4})|^2).
\end{align} 
From this, we can also obtain $\Re[p]=\pm\sqrt{|p|^2-\Im[p]^2}$, where the sign is, however, unknown.

To resolve the sign information, we consider the information gained from $|y(\theta_1, \theta_2)|^2$. 
Expanding Eq.~\eqref{eq:y} with $c_i = \cos(\theta_i)$ and $s_i = \sin(\theta_i)$ yields
\begin{align}
     y(\theta_1, \theta_2) &= c_1c_2r + ic_2s_1p_1 +ic_1s_2p_2 - s_1s_2p\label{eq:y_expand}.
\end{align}
\begin{widetext}
The full expansion of the amplitude $|y(\theta_1,\theta_2)|^2$ is given as:
\begin{align}
|y(\theta_1,\theta_2)|^2
&= c_1^2 c_2^2 r^2
   + c_2^2 s_1^2 |p_1|^2
   + c_1^2 s_2^2 |p_2|^2
   + s_1^2 s_2^2 |p|^2
   + 2 c_1 c_2 s_1 s_2 \,\Re[p_1 p_2^*] \nonumber\\[4pt]
&\quad- 2 c_1 c_2 s_1 s_2 \, r\,\Re[p] 
   - 2 c_1 c_2^2 s_1 \, r\,\Im[p_1]
   - 2 c_1^2 c_2 s_2 \, r\,\Im[p_2] 
   - 2 c_2 s_1^2 s_2 \,\Im[p_{}p_1^*]
   - 2 c_1 s_1 s_2^2 \,\Im[p_{}p_2^*].
\end{align}
We first consider the following four sign variants of a single angle $\theta$, using the shorthand $q_\pm = p_1 \pm p_2$:
\begin{align}
    |y_{+,+}|^2=|y(+\theta,+\theta)|^2 &= c^4 r^2+c^2s^2|q_+|^2+s^4|p|^2 
    -2c^2s^2r\,\Re[p] 
    - 2c^3sr\, \Im[q_+] 
    - 2cs^3\,\Im[p_{}q_+^*],\\
    |y_{-,-}|^2=|y(-\theta, -\theta)|^2 &= c^4 r^2+c^2s^2|q_+|^2+s^4|p|^2 
    - 2c^2s^2r\,\Re[p] 
    + 2c^3sr\, \Im[q_+] 
    + 2cs^3\,\Im[p_{}q_+^*],\\
    |y_{+,-}|^2=|y(+\theta, -\theta)|^2 &= c^4 r^2+c^2s^2|q_-|^2+s^4|p|^2 
    + 2c^2s^2r\,\Re[p] 
    - 2c^3sr\, \Im[q_-] 
    - 2cs^3\,\Im[p_{}q_-^*],\\
    |y_{-,+}|^2=|y(-\theta, +\theta)|^2 &= c^4 r^2+c^2s^2|q_-|^2+s^4|p|^2 
    + 2c^2s^2r\,\Re[p] 
    + 2c^3sr\, \Im[q_-] 
    + 2cs^3\,\Im[p_{}q_-^*].
\end{align}
\end{widetext}
By combining the measurement results and choosing {$\theta=\tfrac{\pi}{4}$}, we obtain the relation
\begin{align}\label{eq:K0}
    K_0&=\tfrac{1}{2}(|y_{+,+}|^2+|y_{-,-}|^2 - |y_{+,-}|^2 - |y_{-,+}|^2)\big|_{\theta=\pi/4} \nonumber\\
    &=\tfrac{1}{4}(|q_+|^2-|q_-|^2-4r\,\Re[p])\nonumber\\
    &=\Re[p_1p_2^*]-r\,\Re[p]\nonumber\\
    &=\Re[p_1]\,\Re[p_2]+\Im[p_1]\,\Im[p_2] - r\,\Re[p].
\end{align}
The imaginary parts $\Im[p_1]$ and $\Im[p_2]$ can be computed by measuring $|y(\tfrac{\pi}{4},0)|^2$ with $|p_1|^2=|y(\tfrac{\pi}{2},0)|^2$, and $|y(0,\tfrac{\pi}{4})|^2$ with $|p_2|^2=|y(0,\tfrac{\pi}{2})|^2$, respectively; similarly to how $\Im[p]$ is obtained.
Therefore, $\Im[p]$, $\Im[p_1]$, and $\Im[p_2]$ are fully known, while $\Re[p]$, $\Re[p_1]$, and $\Re[p_2]$ are known up to a sign.
Rearranging Eq.~\ref{eq:K0} and squaring yields
\begin{align}
    (\underbrace{K_0-\Im[p_1]\,\Im[p_2]}_{=M_0}+r\,\Re[p])^2 = \Re[p_1]^2\Re[p_2]^2.
\end{align}
Therefore, $\Re[p]$ is fully determined as
\begin{align}
    \Re[p]=\frac{\Re[p_1]^2\Re[p_2]^2 - M_0^2 - r^2\Re[p]^2}{2r\,M_0},
\end{align}
unless $M_0=0$.
In this case, we may choose another combination of the amplitudes $|y_{\pm, \pm}|^2$.
Terms with $p_1$ can be isolated by combining as
\begin{align}\label{eq:K1}
    K_1&=\tfrac{1}{2}(|y_{+,+}|^2-|y_{-,-}|^2 - |y_{-,+}|^2+|y_{+,-}|^2)\big|_{\theta=\pi/4} \nonumber\\
    &= -\tfrac{r}{2}(\Im[q_+]+\Im[q_-])
    - \tfrac{1}{2}(\Im[p_{}q_-^*]+\Im[p_{}q_+^*])\nonumber\\
    &= -(r\,\Im[p_1] + \Im[p_{}p_1^*])\nonumber\\
    &= -(r\,\Im[p_1] + \Im[p]\,\Re[p_1]-\Re[p]\,\Im[p_1]);
\end{align}
and similarly for $p_2$:
\begin{align}\label{eq:K2}
    K_2&=\tfrac{1}{2}(|y_{+,+}|^2-|y_{-,-}|^2 + |y_{-,+}|^2-|y_{+,-}|^2)\big|_{\theta=\pi/4} \nonumber\\
    &= \tfrac{r}{2}(\Im[q_-]-\Im[q_+])
    + \tfrac{1}{2}(\Im[p_{}q_-^*]-\Im[p_{}q_+^*])\nonumber\\
    &= -(r\,\Im[p_2] + \Im[p_{}p_2^*])\nonumber\\
    &= -(r\,\Im[p_2] + \Im[p]\,\Re[p_2]-\Re[p]\,\Im[p_2]).
\end{align}
Eq.~\eqref{eq:K1} or Eq.~\eqref{eq:K2} alone are also sufficient to fully determine $\Re[p]$ unless $M_1=K_1 + r\,\Im[p_1] = 0$ or $M_2=K_2 + r\,\Im[p_2] = 0$, respectively.

We comment on the edge case, where $M_0=M_1=M_2=0$.
If $M_i=0\,\forall i$ and $\Im[p]\neq0$, we can insert Eq.~\ref{eq:K1} and Eq.~\ref{eq:K2} into Eq.~\ref{eq:K0}, which determines
\begin{align}
    \Re[p]=\Re[p]^2\,\Im[p_1]\,\Im[p_2]/(r\,\Im[p]^2).
\end{align}
If $M_i=0\,\forall i$ and $\Im[p]=0$, but either $\Im[p_1]\neq0$ or $\Im[p_1]\neq0$, then Eq.~\eqref{eq:K1} or Eq.~\eqref{eq:K2} directly provide
\begin{align}
    \Re[p]=(K_i - r\,\Im[p_i])/\Im[p_i].
\end{align}
If, however, $M_i=0\,\forall i$ and also $\Im[p]=\Im[p_1]=\Im[p^2]=0$, then we cannot resolve the sign of $\Re[p]$ from the equations \eqref{eq:K0}, \eqref{eq:K1}, and \eqref{eq:K2}.

Except for this case, we can obtain $p=\langle\psi|PU|\psi\rangle$ from the measured amplitudes $|x(\theta)|^2$ and $|y(\theta_1, \theta_2)|^2$ together with the known Loschmidt echo $r=\langle\psi|U|\psi\rangle$. 
The edge case, which we expect to be rare in practice, may be overcome by choosing a different pair of operators $P_1$ and $P_2$.

For a concrete application, we may consider a phase gate where $P=Z_1\otimes Z_2$.
Then, a simple decomposition such that $P_1P_2=P$ is found by choosing $P_1=Z_1$ and $P_2=Z_2$. 
For single-qubit $P$, a construction via the two-qubit gates is possible: $P=Z_1\otimes \mathbb{I}$, $P_1=Z_1\otimes Z_2$ and $P_2=\mathbb{I}\otimes Z_2$.
Similar decompositions can be found for all Pauli-strings.
This technique completely avoids auxiliary qubits, which is particularly beneficial when the qubit connectivity is limited. 
We leave a more detailed analysis and comparison to other methods for future work.

\section{Details on the numerical simulation} \label{app:sec:Ising} 

We provide additional information on the numerical case study shown in Fig.~\ref{fig:Fig3} and Fig.~\ref{fig:Fig4}.
The one-dimensional Ising Hamiltonian is decomposed into odd and even terms $H_\text{Ising}=H_\text{odd}+H_\text{even}$ using a second-order Trotter-Suzuki decomposition~\cite{hatano_finding_2005}.
The sequence of gates that we consider is the one described above in Sec.~\ref{app:cutting}.
Whenever possible, we merge consecutive two-qubit gates.

When considering hardware noise, we assume depolarizing noise on all multi-qubit gates, i.e.~on all circuit gates implementing the Ising model dynamics and on the controlled gate, which is a three-qubit operation.
Single qubit operations, which appear in the product state preparation and unpreparation and in the basis transformations on the auxiliary qubit, are simulated without noise, as current hardware can implement them with much higher fidelity than entangling gates.
The $k$-qubit depolarizing channel acting on a state $\rho$ is defined as
\begin{align}
    \mathcal{D}^{(k)}_{\gamma}[\rho]
    =
    (1-\gamma)\rho + \frac{\gamma}{4^k-1}\sum_{P\in\mathcal{P}_k^{\times}}
    P\,\rho\,P,
\end{align}
where $\gamma\in[0,1]$ is the depolarizing noise strength and  
$\mathcal{P}_k^{\times}=\{\mathbb{I},\sigma^x,\sigma^y,\sigma^z\}^{\otimes k}\setminus\{\mathbb{I}^{\otimes k}\}$ denotes the set of all non-identity Pauli strings on $k$ qubits. 
Thus, with total probability $\gamma$ one of the $4^k-1$ non-identity Pauli strings is applied, each occurring with equal likelihood.
We choose a noise strength $\gamma_2=2\times10^{-3}$ in the simulation for the two-qubit gates, corresponding to current hardware~\cite{acharya_quantum_2025}. 
For the controlled two-qubit gate, we consider a depolarizing noise strength of $\gamma_3=6\times10^{-3}$, approximately three times the two-qubit noise level. 
This choice is motivated by the decomposition of a double-controlled phase gate as three controlled phase gates and single qubit rotations.
For simplicity, we assume that connectivity with the auxiliary qubit incurs no additional routing or swap-network cost.
The numerical simulation is implemented using Cirq~\cite{cirq_developers_cirq_2024, quantum_ai_team_and_collaborators_qsim_2020}.

\begin{figure}[t]
    \centering
    \includegraphics[width=1.0\linewidth]{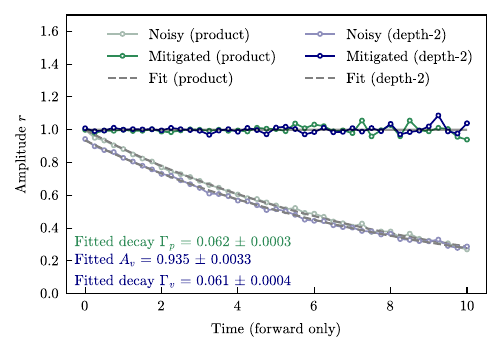}
    \caption{We consider the Ising model $H(\lambda=0.5) = 0.75 \sum_j \sigma_j^z\sigma^z_{j+1} + \sum_j \sigma_j^x + 0.5 \sum_j \sigma_j^z $ on 19 spins with a depolarizing noise strength $\gamma = 2\times 10^{-3}$ for multi-qubit gates. Noise is also applied during state preparation affecting the variational ansatz, hence at $t=0$ the amplitude $r$ deviates from 1. We fit this deviation as $A_v$.
    For each time point, we average over 2000 samples.}
    \label{fig:App_miti}
\end{figure}

\begin{figure}[t]
    \centering
    \includegraphics[width=1.0\linewidth]{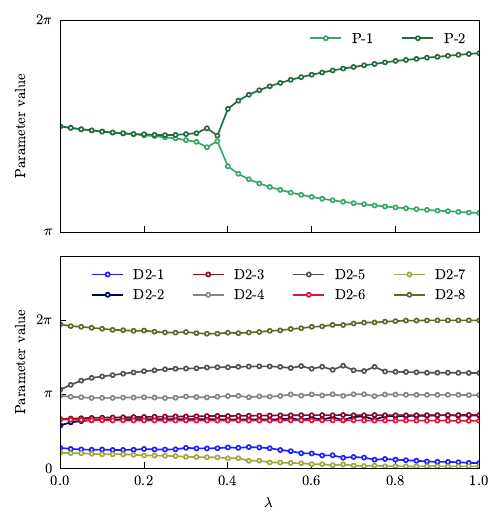}
    \caption{Optimized parameters for the variational ansatz states.
    The product state ansatz (upper panel) has two parameters (P-1 for odd and P-2 for even sites, respectively) and the depth-2 ansatz (lower panel) has eight variational parameters. 
    The first two parameters (D2-1 and D2-2) correspond to the first layer of single qubit rotations $U_1$, D2-3 is the duration of the $U_2$ entangling layer, etc., as described in the Appendix~\ref{app:sec:Ising}.}
    \label{fig:App_params}
\end{figure}

We include the data for error mitigation in the numerical benchmarks of the Ising model. 
As described in \cite{yang_simulating_2023,yang_phase-sensitive_2024}, we evolve forward and backward in time, such that a noiseless circuit returns the initial state with perfect fidelity.
Hardware imperfections introduce deviations from this trajectory, resulting in an exponential decay in the survival probability. 
The decay can be seen in Fig.~\ref{fig:App_miti}, where we fit an exponential function
$r(t) = A\exp(-\Gamma t)$ for both the product state and the depth-2 ansatz.
Note that the depth-2 ansatz has an initial amplitude $A_v<1$ at time $t=0$, as the state preparation is subject to noise.
We observe very good agreement with an exponential fit.

As variational ansätze we consider a product state and a depth-2 ansatz.
The product state has two parameters, one for the odd and one for the even sites,
\begin{align}
    \ket{p(\theta_\text{o}, \theta_\text{e})} = \ket{\phi(\theta_\text{o})} \otimes \ket{\phi(\theta_\text{e})}  \otimes \cdots \otimes \ket{\phi(\theta_\text{o})},
\end{align}
where the state on each site is prepared by a $\sigma^y$-rotation as 
\begin{align}
    \ket{\phi(\theta)} &= \exp(-i\sigma^y\theta/2) \ket{0} \nonumber\\
    &=\cos(\theta/2) \ket{0} - \sin(\theta/2) \ket{1}.
\end{align}

For the depth-2 ansatz, we combine a layer of single-qubit rotations 
\begin{align}
    U_1(\theta_{\text{o}},\theta_{\text{e}})=\exp(-i\sigma^y\theta_{\text{o}}/2)\otimes \exp(-i\sigma^y\theta_\text{e}/2)\otimes\dots
\end{align}
with an entangling layer, which is a time-dependent evolution with the classical part of the Ising Hamiltonian:
\begin{align}
    U_2(\beta) = \exp[-iH_\text{Ising}(1.5,0,1)\beta ].
\end{align}
The depth-2 ansatz consists of alternating three $U_1$ and two $U_2$ layers, with eight parameters in total.
We include the optimized parameters in Fig.~\ref{fig:App_params}.

For the Gaussian filter $P_\delta(E)$ to compute the local density of states $ D(E) = \langle\psi|P_\delta(E)|\psi\rangle$, we replace the continuous integral by a sum over $2R+1$ equally spaced times:
\begin{align}
    D(E) \approx \sum_{m=-R}^{R} c_m e^{it_mE}g(t_m).
\end{align}
Here, the spacing between the times is $\Delta t=t_{\max}/R$ and the Loschmidt echoes need to be computed for the discrete times $t_m=m\Delta t$.
The coefficients $c_m=\Delta t\frac{\delta}{\sqrt{2\pi}}\exp(-t_m^2\delta^2/2)$ are computed classically.

\section{Numerical comparison with the global Hadamard test} \label{app:Hadamard_compare}

\begin{figure*}[t]
    \centering
    \includegraphics[width=1.0\linewidth]{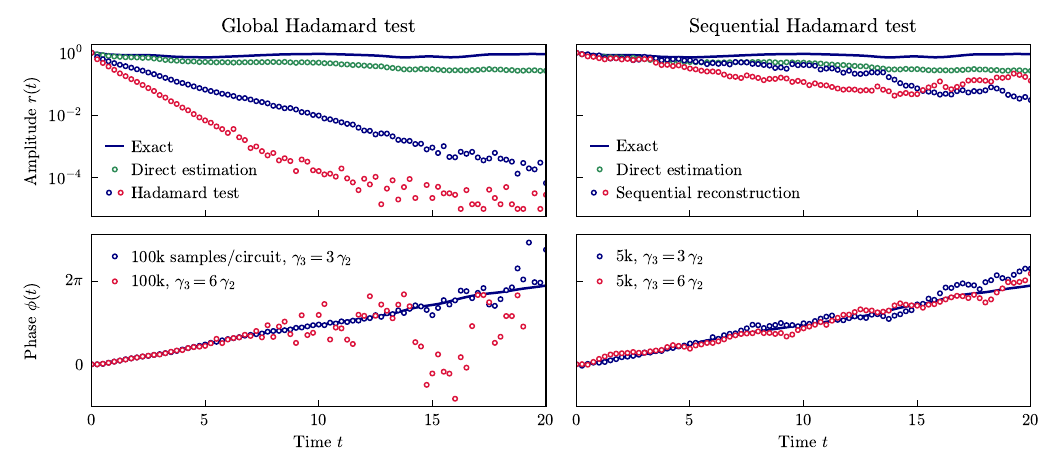}
    \caption{Numerical comparison of the global (standard) Hadamard test (left) and the sequential Hadamard test (right). 
    The upper panels show the amplitude $r(t)$ and the lower panels the phase $\phi(t)$ of the Loschmidt echo (after subtracting the phase contribution of the initial state's mean energy).
    Solid lines denote exact values, while markers represent estimated values. 
    The green markers (direct estimation) show the amplitude obtained without any controlled gates from 5,000 samples per data point.
    Blue and red markers consider two noise strengths $\gamma_3$ for the three-qubit gates, with a two-qubit depolarizing rate $\gamma_2=2\times10^{-3}$. 
    Each circuit is repeated 100,000 times for the global Hadamard test and 5,000 times for the sequential Hadamard test, such that the total sample budget for the two methods is comparable.
    Higher $\gamma_3$ accelerates the decay of the amplitude, rendering the computation of the phase unreliable at late times for the global Hadamard test.
    }
    \label{fig:App_Hadcompare}
\end{figure*}

We perform a numerical benchmark of the sequential Hadamard test against the global Hadamard test. The model and the initial state are the same as in Fig.~\ref{fig:Fig4}.
In the global Hadamard test, every two-qubit gate is promoted to a controlled gate. 
The implementation of a controlled gate depends on the available gate set of the quantum computer; however, it will generally have a lower fidelity than its uncontrolled gate.

Limited qubit connectivity yields additional overhead in the global Hadamard test when the controlling qubit is not a direct neighbor of the controlled gate.
The precise number of additional swapping or shuttling operations depends on the device architecture and the Hamiltonian. 
We therefore intentionally exclude such swaps and assume all-to-all connectivity, which strongly favors the global Hadamard test.

Despite excluding swaps, we can clearly observe the existence of a favorable regime for the sequential Hadamard test in our simulations. 
We consider a two-qubit-gate depolarizing noise rate of $\gamma_2=2\times10^{-3}$ and two different noise rates $\gamma_3=3\,\gamma_2$ and $\gamma_3=6\,\gamma_2$ for the controlled two-qubit gate.
For a fair comparison, we allow 20 times as many samples per circuit in the global Hadamard test, such that the total sample count of both methods is approximately the same.

We include the numerical results in Fig.~\ref{fig:App_Hadcompare}. 
The amplitude data are included only as a reference; in an experiment, one directly estimates the amplitude without controlled operations (green markers).
As expected, the decay of the amplitude depends on the noise strength $\gamma_3$, with larger noise leading to a faster decay. 
For the higher noise rate $\gamma_3=6\,\gamma_2$, the phase information in the global Hadamard test is completely scrambled for times $t \gtrsim 10$, corresponding to the regime where the shot noise in the amplitude becomes larger than the amplitude itself.
For the sequential Hadamard test, we find that the fluctuations in the phase are significantly smaller at later times than in the global Hadamard test for realistic parameters.
Hence, there is a relevant regime where the sequential Hadamard test allows one to obtain the complex Loschmidt echo at larger times than with previous methods.

\section{Effect of Loschmidt echo amplitude damping on the local density of states} \label{app:broadening}

We note that if noise can be modeled as exponential damping of the amplitude $r(t)$, and we do not mitigate the error, then the hardware errors lead to an effective broadening of the reconstructed local density of states (LDOS).
For concreteness, we may consider that we obtain damped Loschmidt echoes 
\begin{align}
    \tilde{g}(t)=g(t)\exp(-\Gamma t).
\end{align} 
The LDOS is $D(E) = \langle\psi|\delta(H-E)|\psi\rangle$  with $\delta(H-E)$ a $\delta$-function centered at an energy $E$. 
Therefore, we obtain 
\begin{align}
    D(E) = \int_{-\infty}^{\infty}\di t\;  \tilde{g}(t) e^{iEt} =\int_{-\infty}^{\infty}\di t\; g(t) e^{i(E-i\Gamma)t},
\end{align}
implying that the exponential damping of the amplitudes yields a convolution of Gaussians and Lorentzians, where the Lorentzians correspond to an additional line-broadening.

\section{Considerations for the Rydberg Hamiltonian} \label{app:Rydberg}

We consider Ising-type models, which can be split into a classical and a quantum part $H = H_c + H_q$, with $H_q\propto\sum_i \sigma_i^x$. 
Then, a classical initial state is an eigenstate of $H_c$, and at a first-order Trotter decomposition, the ITE gate only yields a global prefactor $\exp(-\zeta_c\tau)$:
\begin{align}
    e^{-H \tau} \ket{\psi} \approx e^{-H_q\tau } e^{-H_c\tau } \ket{\psi} = e^{-\zeta_c\tau } e^{-H_q\tau } \ket{\psi}.
\end{align}
This simplification applies, in particular, to the many-body Rydberg Hamiltonian defined on a graph $\mathcal{G}=(\mathcal{V},\mathcal{E})$ with vertices $\mathcal{V}$ and edges $\mathcal{E}$ as:
\begin{align}
    H_\text{Ryd} = \sum_{(i,j) \in \mathcal{E}} \frac{V_0n_i n_j}{|r_i-r_j|^6}  - {\Delta} \sum_{i \in \mathcal{V}} n_i + \frac{\Omega}{2} \sum_{i \in \mathcal{V}} \sigma^x
\end{align}
with $n_i=|1\rangle\langle1|_i$.
The model can be natively implemented on neutral atom arrays, which have emerged in recent years as a powerful hardware platform for quantum computation allowing for both analog simulation and high-fidelity digital gates~\cite{bluvstein_quantum_2022}. 
Realizing the ITE for the quantum part $H_q$ is particularly straightforward in this case, as $\exp(-H_q\tau )$ can be implemented with local $\sigma^y$-rotations. 

Interestingly, in the strong-interaction limit of the so-called PXP model~\cite{lesanovsky_interacting_2012}, short ITE can be realized for arbitrary product states with local $\sigma^y$-rotations, instead of only classical product states.
The PXP model is obtained in the strong-interaction limit of the Rydberg Hamiltonian, neglecting contributions from the long-range interaction tails. We rewrite the local blockade constraint into a global constraint as 
\begin{align} \label{eq:PXP}
    H &= \frac{\Omega}{2}\sum_j P^{}_{j-1}\sigma_j^x P^{}_{j+1} - \Delta \sum_j n_j \nonumber\\
    &= \mathcal{P} \bigg( \sum_j \frac{\Omega}{2} \sigma_j^x - \Delta \sum_j n_j \bigg) \mathcal{P}\nonumber\\
    &= \mathcal{P} \bigg(\sum_j h_j \bigg) \mathcal{P}. 
\end{align}
For the PXP model, any initial state in the subspace can be used for implementing ITE with single site rotations.
We consider an initial state ansatz parametrized by an angle $\alpha$. 
For each spin on sublattice A, we choose the initial spin as $\ket{\phi(\alpha)} = \cos(\alpha/2) \ket{0} + \sin(\alpha/2) \ket{1}$ and the full initial state is
\begin{align}
    \ket{\psi(\alpha)} = \ket{\phi(\alpha)} \otimes \ket{0} \otimes \ket{\phi(\alpha)} \otimes \cdots \otimes \ket{\phi(\alpha)}.
\end{align}
The Hamiltonian for a single spin can be written as
\begin{align}
    h &= \begin{pmatrix}
        0 & \Omega/2\\
        \Omega/2 & -\Delta
    \end{pmatrix}
    =\frac{\Omega}{2}\sigma^x - \Delta n \nonumber\\
    &= \Omega_\text{eff} [\sin(\theta) \sigma^x - 2\cos(\theta) n]
\end{align}
where the effective Rabi frequency $\Omega_\text{eff}=\sqrt{\Delta^2 + \Omega^2}/2$ and $\tan(\theta)=\Omega/\Delta$, i.e.~$\Omega_\text{eff} \sin(\theta) = \Omega/2$ and $\Omega_\text{eff} \cos(\theta) = \Delta/2$.
Then, we compute that a $\sigma^y$-rotation is sufficient to realize short ITE under this model. We can see this by explicitly computing
\allowdisplaybreaks
\begin{align}
& (h-\langle h \rangle) \ket{\phi(\alpha)} \nonumber\\
=\,& 
{\renewcommand\arraystretch{1.5}
 \begin{pmatrix}
     -\frac{\Omega}{2} \sin(\alpha/2) \cos(\alpha)  + \frac{\Delta}{2}\sin(\alpha/2)\sin(\alpha) \\ 
      \frac{\Omega}{2} \cos(\alpha/2)\cos(\alpha) - \frac{\Delta}{2} \cos(\alpha/2)\sin(\alpha) 
 \end{pmatrix}} \nonumber\\
 =\,& \frac{1}{2} [-\Omega \cos(\alpha) + \Delta\sin(\alpha)] 
\begin{pmatrix}
    \sin(\alpha/2)\\
    -\cos(\alpha/2)
\end{pmatrix} \nonumber\\
=\,& \Omega_\text{eff} \sin(\alpha-\theta) 
\begin{pmatrix}
    0 & 1\\
    -1 & 0
\end{pmatrix}
\ket{\phi(\alpha)} \nonumber\\
=\,& i_{} a_{} \sigma^y \ket{\phi(\alpha)},
\end{align}
where we define $a(\alpha,\theta) = \Omega_\text{eff} \sin(\alpha-\theta)$.
Hence, in the PXP model, imaginary time evolution for initial classical states can be implemented with local $\sigma_y$-rotations.

\bibliography{biblio}

\end{document}